\documentclass[10pt,twocolumn,letterpaper]{article}

\usepackage{iccv}              % To produce the CAMERA-READY version
\definecolor{iccvblue}{rgb}{0.21,0.49,0.74}
\usepackage[pagebackref,breaklinks,colorlinks,allcolors=iccvblue]{hyperref}
\usepackage{placeins} 

\def\paperID{9222} % *** Enter the Paper ID here
\def\confName{ICCV}
\def\confYear{2025}

\title{Aberration Correcting Vision Transformers for High-Fidelity Metalens Imaging}

\newcommand\CorrespondingAuthorMark{\textsuperscript{†}}

\author{Byeonghyeon Lee$^1$ \and
Youbin Kim$^2$ \and
Yongjae Jo$^2$ \and
Hyunsu Kim$^2$ \and
Hyemi Park$^2$ \and
Yangkyu Kim$^2$ \and
Debabrata Mandal$^3$ \and
Praneeth Chakravarthula$^3$\CorrespondingAuthorMark \and
Inki Kim$^2$\CorrespondingAuthorMark \and
Eunbyung Park$^1$\CorrespondingAuthorMark \\
$^1$ Yonsei University \\
$^2$ Sungkyunkwan University \\
$^3$ University of North Carolina at Chapel Hill}

\begin{document}
\maketitle
\begingroup
\renewcommand\thefootnote{}
\footnotetext{\textsuperscript{†}Corresponding authors.}
\endgroup
\begin{abstract}
Metalens is an emerging optical system with an irreplaceable merit in that it can be manufactured in ultra-thin and compact sizes, which shows great promise in various applications. Despite its advantage in miniaturization, its practicality is constrained by spatially varying aberrations and distortions, which significantly degrade the image quality. Several previous arts have attempted to address different types of aberrations, yet most of them are mainly designed for the traditional bulky lens and ineffective to remedy harsh aberrations of the metalens. While there have existed aberration correction methods specifically for metalens, they still fall short of restoration quality. In this work, we propose a novel aberration correction framework for metalens-captured images, harnessing Vision Transformers (ViT) that have the potential to restore metalens images with non-uniform aberrations. Specifically, we devise a Multiple Adaptive Filters Guidance (MAFG), where multiple Wiener filters enrich the degraded input images with various noise-detail balances and a cross-attention module reweights the features considering the different degrees of aberrations. In addition, we introduce a Spatial and Transposed self-Attention Fusion (STAF) module, which aggregates features from spatial self-attention and transposed self-attention modules to further ameliorate aberration correction. We conduct extensive experiments, including correcting aberrated images and videos, and clean 3D reconstruction. The proposed method outperforms the previous arts by a significant margin. We further fabricate a metalens and verify the practicality of our method by restoring the images captured with the manufactured metalens. Code and pre-trained models are available at \href{https://benhenryl.github.io/Metalens-Transformer}{https://benhenryl.github.io/Metalens-Transformer}.
\end{abstract}

\section{Introduction}
\label{sec:intro}

\begin{figure}[t]
\begin{center}
\includegraphics[width=1.0\linewidth]{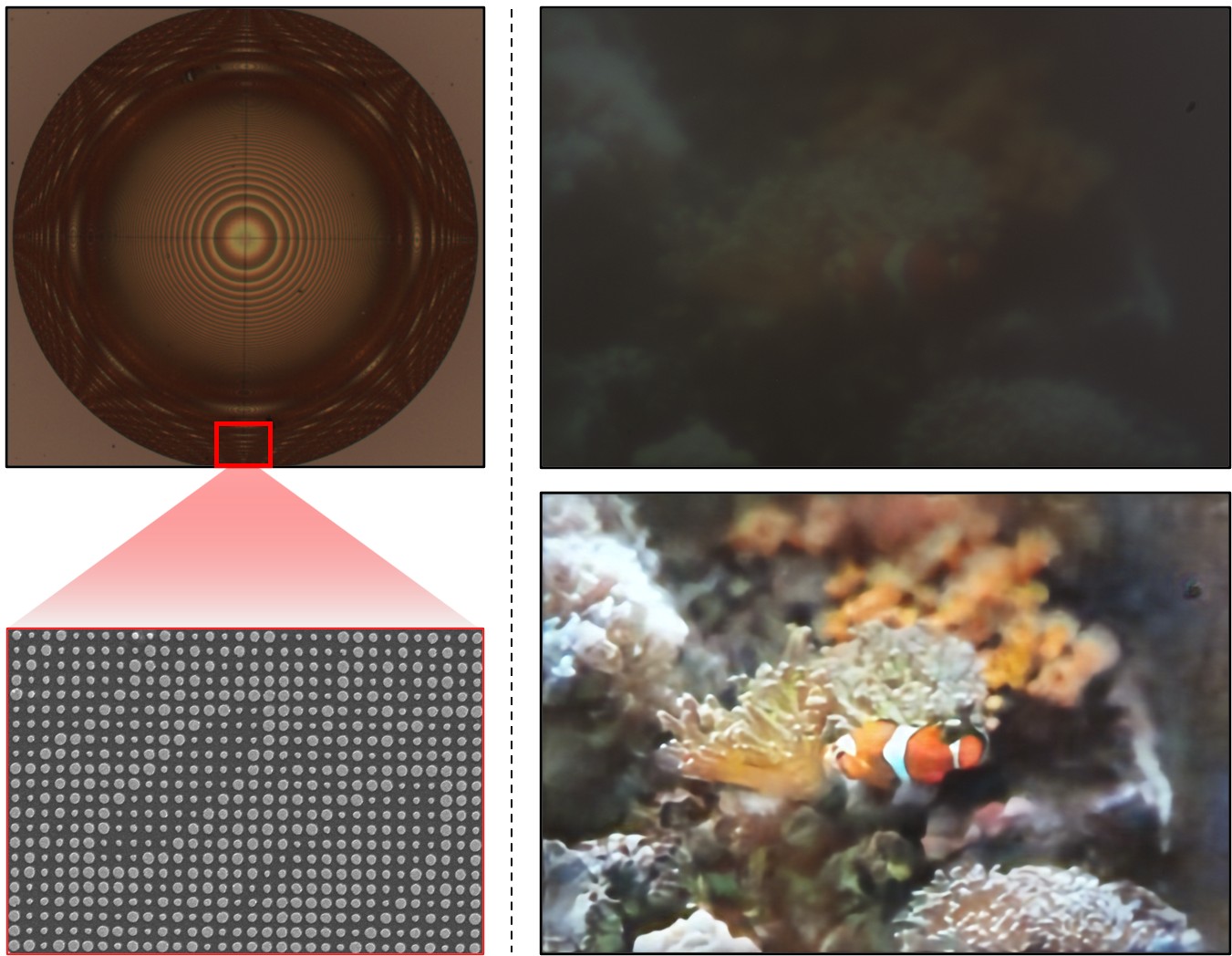}
\end{center}
\vspace{-0.5cm}
   % \caption{Qualitative results on correcting the aberration of the images captured with the fabricated metalens.}
   \caption{Left: The fabricated metalens. Right: An image captured with the manufactured metalens (top) and restored image with the proposed method (bottom).}
\vspace{-0.5cm}
\label{fig:quali_real}
\end{figure}

Metalens is an advanced optical device that leverages nanoscale engineering to manipulate light. Unlike traditional lenses relying on glass or plastic surfaces, the metalens harnesses a metasurface, a thin optical element consisting of nanostructures modulating optical properties such as phase and amplitude. It can be fabricated in ultra-thin and compact sizes, making it an ideal option for various domains, including the medical industries, consumer electronics, and augmented/virtual reality (AR/VR) \cite{zhang2020metasurfaces, stoja2021improving, yoon2021printable, gu2023reconfigurable, park2019all, li2021meta,kim2021dielectric, li2022ultracompact}. Despite its potential for miniaturization, existing metalenses have faced critical challenges, such as aberrations and distortions in the captured images, hampering their broad uses across diverse applications.

Aberrations in metalens occur due to practical constraints and the physical properties of lens materials. Wavelength-dependent nanostructures of the metasurface make different wavelengths of lights experience different phase shifts, which results in them being focused at different points, leading to chromatic aberration. Furthermore, its spherical surface is susceptible to spherical aberrations where rays transmitting through the edges and center of the lens focus at different locations. Even worse, these aberrations are not uniform, rather varying across the image field, posing a more intricate problem of correcting them.

%Several previous works~\cite{chang2012correction,eboli2022fast,gong2024physics} have attempted to correct these aberrations, primarily addressing them for general commercial cameras and not effective enough to mitigate intense aberrations of metalens. On the other hand, \citet{tseng2021neural} and~\citet{chakravarthula2023thin} devised aberration correcting methods for the metalens. \citet{tseng2021neural} adopted convolutional neural networks (CNNs) and~\citet{chakravarthula2023thin} employed a diffusion model~\cite{ho2020denoising,song2020denoising} for restoration. While promising, these methods have yet to reach the desired restoration quality.

Several previous works~\cite{chang2012correction,eboli2022fast,gong2024physics} have attempted to correct these aberrations, primarily focusing on conventional camera lenses. While effective in some cases, these approaches fall short in addressing the unique challenges posed by metalens-induced aberrations. More recently, \citet{tseng2021neural} and~\citet{chakravarthula2023thin} have proposed deep learning-based restoration methods for metalenses. \citet{tseng2021neural} adopted convolutional neural networks (CNNs) and~\citet{chakravarthula2023thin} employed a diffusion model~\cite{ho2020denoising,song2020denoising} to correct distortions. While promising, they have yet to reach the desired restoration quality.

% Besides the aberration corrections, numerous studies have tackled other image degradations such as blur, noise, raindrops, haze,~\etc. A line of research introduced~\cite{kawar2022denoising,murata2023gibbsddrm,kupyn2018deblurgan,kupyn2019deblurgan} generative models~\cite{goodfellow2020generative,song2020denoising,ho2020denoising} to produce clean images from degraded images, and other arts~\cite{Liu2019MWCNN, zhang2017learning, Mao2016ImageRU, He_2019_CVPR, lecouat2020fully,Zamir2021MPRNet} adopted CNNs. Another line of works~\cite{zamir2022restormer,swinir,xrestormer,dat,ipt} harnessed Vision Transformer (ViT)~\cite{vit} for image restoration. Although they have shown prominent results, straightforwardly applying them to metalens-induced aberration correction is undesirable. This is because metalens can manipulate multiple optical properties, resulting in complex aberration patterns that vary in response to different wavelengths, which lead to severe chromatic aberrations. Moreover, aberrations in metalens images can amplify noise and distort intensity variations across channels. 
% These aberrations are more complicated than the degradations encountered in conventional image restoration tasks, and those approaches devised for general image restoration lack in capacity to adapt to these distortions.
Besides the aberration corrections, extensive research has addressed other image degradations such as blur, noise, raindrops, haze,~\etc. Generative models~\cite{kawar2022denoising,murata2023gibbsddrm,kupyn2018deblurgan,kupyn2019deblurgan}, CNNs~\cite{Liu2019MWCNN, zhang2017learning, Mao2016ImageRU, He_2019_CVPR, lecouat2020fully,Zamir2021MPRNet}, or Vision Transformer (ViT)~\cite{zamir2022restormer,swinir,xrestormer,dat,ipt} have all been employed for image restoration, demonstrating strong performance. However, directly applying these methods to metalens-induced aberration correction is ineffective due to the aberration patterns of metalens that vary in response to different wavelengths, making them significantly more complex and challenging to correct using standard restoration techniques.
%Although they have shown prominent results, straightforwardly applying them to metalens-induced aberration correction is undesirable due to the aberration patterns of metalens that vary in response to different wavelengths, and are more complicated than the degradations encountered in conventional image restoration tasks.

Among various frameworks, ViT-based approaches have a strong potential for metalens image restoration. ViT can effectively capture global context by leveraging the self-attention mechanism,  allowing each pixel in an image to communicate with all others. This global receptive field is particularly advantageous for metalens restoration, where understanding the overall structure helps suppress noise and artifacts while preserving fine details. Moreover, self-attention modules enable dynamic weighting of features, encouraging them to learn spatially varying corrections, which is essential in metalens image restoration since the severity of aberration is different across the image. Despite these advantages, ViT-based methods remain largely unexplored for metalens restoration.

%where understanding the overall structure plays a key role in removing noise or artifacts while preserving intricate details. Moreover, self-attention modules enable dynamic weighting of features, encouraging them to learn spatially varying corrections, which is essential in metalens image restoration since the severity of aberration is different across the image. CNNs, on the other hand, can interact only with neighboring pixels and apply the same kernel to every region in the image, restricting modeling spatially varying aberrations.  While ViT has such competence in restoring metalens images, it remains largely unexplored. 

In this work, we propose a novel aberration correction Transformer specifically designed for metalens image restoration. First, we suggest a Multiple Adaptive Filters Guidance (MAFG) module, which leverages a set of Wiener filters~\cite{wiener} alongside a cross-attention mechanism to dynamically weight features for more effective aberration correction. MAFG offers two key advantages. First, it is well-suited for handling spatially varying aberrations. MAFG produces features that can consider different degrees of aberration by reweighting multiple features generated with different filters via cross-attention. Second, by employing filters with varying noise suppression parameters, MAFG enhances robustness against unknown aberration distributions, making it adaptable to a wide range of imaging conditions.

%It motivated us to propose an aberration correction Transformer tailored for correcting the aberration of metalens-captured images. Particularly, we design a Multiple Adaptive Filters Guidance (MAFG) that employs several Wiener filters~\cite{wiener} and a cross-attention module for weighting features.
%The advantages of MAFG are threefold: first, it can better cope with the spatially varying aberrations. 
%MAFG produces features that can consider different degrees of aberration by reweighting multiple features generated with different filters via cross-attention.
%Secondly, the filters of MAFG are defined with varied noise suppression parameters, enabling MAFG to deal with unknown distributions. Lastly, we set these parameters based on the image intensity, allowing flexible handling of diverse images.
% across a broad range of images. 

Furthermore, we propose a Spatial and Transposed self-Attention Fusion (STAF) module, which integrates both spatial and transposed self-attention to enhance feature aggregation while considering the distinct roles of the encoder and decoder. In STAF, spatial and transposed self-attention operate in parallel, with their features fused to guide the encoder in capturing global context and the decoder in refining fine details. We also encourage attention modules in the decoder to better capture intricate details, further improving restoration quality.

We conducted extensive experiments across various imaging scenarios to validate the effectiveness of the proposed method. On image dataset with spatially varying aberration, it demonstrated notable performance improvements. We further applied the proposed techniques to video restoration and 3D reconstruction tasks to confirm their superiority. Finally, we fabricated metalens using the open-source PSF of a metalens~\cite{tseng2021neural} and evaluated our method on the captured images. The corrected images demonstrated promising results, revealing new possibilities in metalens imaging.
% cross attention and others
To sum up, our contributions are as follows:

\begin{itemize}
    \item We propose a novel neural architecture for metalens image restoration, achieving state-of-the-art performance with a 6.45 dB improvement in PSNR over the baseline model.
    \item We devise a Multiple Adaptive Filters Guidance (MAFG) that employs several Wiener filters operating adaptively to various images to enrich features of representations.
    \item We design a Spatial and Transposed self-Attention Fusion (STAF) module that applies spatial and transposed self-attention well aligned with encoder-decoder architecture.
    \item We fabricated a metalens with the open-source PSF and constructed a metalens-captured image dataset.
\end{itemize}

\section{Related Works}
\label{sec:related_works}
% XRestormer~\cite{xrestormer} pointed out Restormer
\paragraph{Metalens}

Recent advancements in nanofabrication have enabled the use of ultra-thin metasurfaces consisting of subwavelength scatterers that can independently adjust the amplitude, phase, and polarization of incident wavefronts, allowing for greater design flexibility~\cite{Engelberg2020TheAO,Lin298Dielectric,Mait2020PotentialAppl,froch2024beating}. 
This flexibility has motivated research into applications such as flat meta-optics for imaging~\cite{Colburn2018MetasurfaceOF,Yu2014FlatOW,Aieta2012AberrationFree,Lin2021EndtoEnd,chen2022planar,chakravarthula2023thin,tseng2021neural,froch2024beating}, polarization control~\cite{Arbabi2015DielectricMF}, and holography~\cite{Zheng2015MetasurfaceHR}. 
Notably, work by \citet{tseng2021neural} introduced a differentiable design for meta-optics achieving high image quality within a 0.5 mm aperture but with limitations in field performance beyond 40$^\circ$. 
More recently, \citet{chakravarthula2023thin} introduced a metalens array design that enhances image quality across a full broadband spectrum over a 100$^\circ$ field of view (FoV) without increasing the back focal length, potentially allowing for one-step fabrication directly on the camera sensor coverglass.
However, existing meta-optics still face challenges with chromatic and geometric aberrations, limiting broadband imaging capabilities~\cite{Yu2014FlatOW,Lin298Dielectric,Aieta2015MultiwavelengthAM,Wang2018ABA}. 
Although dispersion engineering methods have mitigated some chromatic aberrations~\cite{Ndao2020OctaveBP,Shrestha2018BroadbandAD,Arbabi2017ControllingTS,khorasaninejad2017achromatic,wang2017broadband}, they remain restricted to very small apertures~\cite{Presutti2020FocusingLimits}. 
In this work, we introduce a framework to compensate for aberrations and overcome the challenges faced by such meta-optics imaging cameras.

\paragraph{Image Restoration Models}
Image restoration aims to recover high-quality images from degraded observations. Traditionally, Wiener deconvolution \cite{wiener} has been used to recover images by solving inverse problems. However, it often struggled with noise and complex degradations, limiting their real-world applications. Recently, deep learning-based methods have shown great success in image restoration tasks in a data-driven manner. Vision Transformer (ViT)~\cite{vit}, in particular,  has achieved remarkable performances in various image restoration tasks, including denoising~\cite{ipt, chen2020pre, Wang_2022_CVPR, fan2022sunet, zamir2022restormer, li2023efficient}, deblurring~\cite{Wang_2022_CVPR, zamir2022restormer, liu2024deblurdinat, kong2022efficientfrequencydomainbasedtransformers, Tsai2022Stripformer}, and super resolution~\cite{chen2023activating, hat, swinir, sun2023safmn, lu2022transformersingleimagesuperresolution, dat}, by exploiting self-attention, capturing global interactions effectively.

Restormer~\cite{zamir2022restormer} stood out with its superior performance, introducing transposed self-attention mechanisms that apply self-attention across the channel of an image, enabling efficient processing of high-resolution images.
% while capturing global contextual information. 
While this approach enhances computational efficiency, it may limit the model’s ability to fully exploit spatial dependencies. To address this, X-Restormer~\cite{xrestormer} incorporated spatial self-attention mechanisms allowing for better capture of spatial relationships.
%However, X-Restormer applies the same attention blocks to both the encoder and decoder, despite their distinct roles in feature representation and reconstruction.
In this work, building upon these advancements, we propose a novel architecture that further enhances restoration quality for metalens images.
%This allowed the model to better capture spatial dependencies, still there is room for improvement in that it deployed attention blocks identically for encoder and decoder though they have different roles when learning the representation.

\begin{figure*}[h]
\begin{center}
\includegraphics[width=1.0\linewidth]{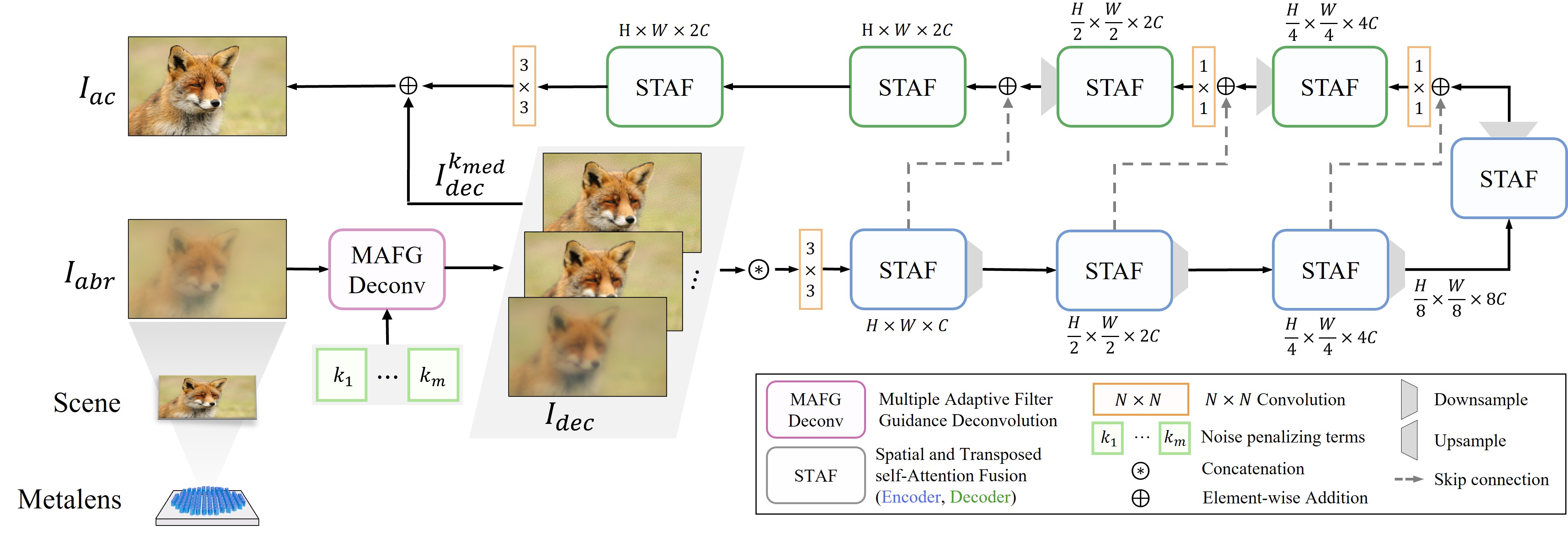}
\end{center}
\vspace{-0.5cm}
   \caption{The overview of our method. It comprises Multiple Adaptive Filters Guidance (MAFG) which produces different representations with various noise-detail balances, and a Spatial and Transposed self-Attention Fusion (STAF) module that aggregates features differently in encoder and decoder.}
\vspace{-0.5cm}
\label{fig:architecture}
\end{figure*}

\paragraph{Aberration Correction Models}
Optical aberration arises from imperfections in optical systems, with chromatic and spherical aberrations being frequently observed in the metalens imaging process. Chromatic aberration, where different wavelengths of light focus at different points, leads to color fringing, and spherical aberration occurs when the rays passing through the edges and center of the lens are focused at different points, resulting in blurriness. Extensive research has been conducted to correct these aberrations, and they can be broadly categorized into non-blind and blind aberration correction. 

Non-blind approaches~\cite{joshi2008psf, kee2011modeling, shih2012image, heide2013high,cho2011handling,dong2021dwdn, zhang2023infwide} require PSF calibration before training the models. Once the PSF is well calibrated, it can be consistently applied across various images, as optical aberrations are independent of the scene content in general \cite{yue2014image, yue2015blind}. 
Especially, DWDN~\cite{dong2021dwdn} leveraged Wiener deconvolution~\cite{wiener} in the feature space for multi-scale refinement and \citet{li2021universal} proposed a PSF-aware network using patch-wise deconvolution with deep priors to cope with spatially varying aberrations.

Blind aberration correction approaches~\cite{rahbar2011blind, schuler2012blind, tang2013does, yue2015blind, eboli2022fast, gong2024physics} correct aberrations by estimating aberration kernels without any prior PSF information.
\citet{eboli2022fast} adopted a small CNN for color correction, and~\citet{gong2024physics} utilized the physical properties of lenses to improve image quality. These blind methods can be applied generally without demanding preprocess (\eg, PSF calibration), but restoration quality may suffer due to uncertainties in accurately estimating the PSF.

Several aberration correction models have been specifically designed for metalens imaging~\cite{tseng2021neural,chakravarthula2023thin}.
\citet{tseng2021neural} employed CNNs and Wiener filters for restoration but faced challenges in preserving fine image details.
This limitation may arise from the restricted receptive field of CNNs.
\citet{chakravarthula2023thin} introduced diffusion models~\cite{ho2020denoising,song2020denoising} with Wiener filters to manage aberrations, yet it can generate unrealistic details due to the stochasticity of the diffusion, compromising the overall image quality.
% \red{yet the reconstructed fine details can differ from the ground-truth in highly ill-posed imaging scenario \cite{yosef2024difuzcam}.} 

%Aberration correction poses unique challenges in the context of metalens imaging, but ViT has not been fully examined to address them. In this paper, we propose leveraging the ViT, adapting attention mechanisms to deal with complex aberration patterns in metalens imaging, to overcome the limitations faced by CNN and diffusion models.
Aberration correction presents unique challenges in metalens imaging, yet the potential of Vision Transformers (ViT) has not been fully explored for this task. In this paper, we propose a novel approach that leverages ViT, developing new architectural modules and adapting its attention mechanisms to effectively handle the complex aberration patterns in metalens imaging.

\section{Methods}
\label{sec:methods}
\cref{fig:architecture} depicts the overall architecture of the proposed method. It consists of Vision Transformer (ViT) as a backbone network for restoration (\cref{sec:method_vit}), Multiple Adaptive Filters Guidance (MAFG) (\cref{sec:method_MAFG}), and Spatial and Transposed self-Attention Fusion (STAF) blocks (\cref{sec:method_STAF}).

\subsection{Preliminary}
\noindent\textbf{Metalens Imaging}
An imaging formulation using a spatially varying point spread function (PSF) can be written in patch-level imaging as follows,
\begin{equation}
\label{eq:imaging}
I_{abr}[h,w] = I_c[h,w] * PSF[h,w] + n[h,w],
\end{equation}
where $[\cdot,\cdot]$ is an indexing operator, $*$ denotes the convolution operator, $h\in\{1,2,...,H\}$ and $w\in\{1,2,...,W\}$ are the indices of $H \times W$ image and PSF patches. $I_{abr}[h,w]$ is the observed image, which can be represented as the convolution of an object $I_c[h,w]$ with a corresponding PSF at patch $[h,w]$. Then the noise $n[h,w]$, such as shutter and natural noise, is added to generate the captured image $I_{abr}$.
% Convolution can be replaced with simple element-wise multiplication when $I_c$ and $PSF$ are in the frequency domains~\cite{fft}. 
The PSF of metalens varies significantly with wavelength because different colors focus at different distances from the metalens. This produces color fringing and a low contrast in the captured image. Such optical degradations vary gradually across the field of view, implying the distortions are similar within a small neighborhood; hence, the imaging can be formulated at a patch level. 

\subsection{Vision Transformer for Metalens Images}
\label{sec:method_vit}

As discussed, various ViT-based approaches~\cite{ipt, swinir, Wang_2022_CVPR, liu2024deblurdinat, hat, sun2023safmn, dat, zamir2022restormer} have been developed for restoring images captured by traditional cameras. While ViT has not been fully explored for metalens image restoration, it inherently has promising capabilities to handle the severe aberrations present in metalens images. First, ViT can adaptively adjust its attention weight to focus on different areas of the image, making it particularly suited for handling the spatially varying aberrations of metalens imaging. Furthermore, chromatic aberrations of metalens differ across different wavelengths of light, causing each image channel to be affected differently. 
ViT’s multi-head attention mechanism allows it to model these channel-specific discrepancies, processing each channel effectively. Lastly, metalens-induced aberrations can degrade the images at various scales, affecting both high-frequency details (\eg, color fringes at sharp edges) and low-frequency components (\eg, gradual intensity shifts). ViT’s ability to capture multi-scale features enhances its capacity to correct different types of distortions.
%ViT can model these channel-specific differences, leveraging its multi-head attention to process each channel properly. Lastly, metalens-induced aberrations can degrade the images at various scales, affecting both high-frequency details (\eg, color fringes at sharp edges) and low-frequency components (\eg, gradual intensity shifts). ViT can represent multi-scale features, which can help in correcting different types of distortions. 

Such capabilities motivated us to adopt ViT for metalens image restoration. Nevertheless, applying ViT directly to metalens image restoration is ineffective, as metalenses introduce more complex aberrations and distortions than conventional lenses, demanding a more specialized solution. Therefore, we employ ViT as the backbone network and incorporate novel aberration-correcting modules to better align ViT with the specific challenges of metalens imaging. %Additionally, we train ViT non-blindly, harnessing PSF.

%As discussed earlier, various ViT-based approaches~\cite{} to restore images captured with traditional cameras were proposed. Among them, Restormer~\cite{zamir2022restormer} showed remarkable restoration performance across several tasks. It adopted encoder-decoder architecture with transposed self-attention (TA)~\cite{} and a novel feed-forward network with gating mechanism~\cite{} and depth-wise convolutions~\cite{}. Nevertheless, simply implementing Restormer for metalens image restoration does not yield satisfactory outcomes because metalens shows more complex aberrations and distortions than general image restoration tasks for conventional lenses, so more sophisticated approach is required. Hence we employ it as a backbone network for restoration and develop novel aberration-correcting modules to make ViT well aligned with metalens images. We also train ViT in a non-blind manner, harnessing point spread function (PSF) for training, as the PSF for metalens can be easily obtained via simple calibration, compared to the traditional cameras.

\subsection{Multiple Adaptive Filters Guidance}
\label{sec:method_MAFG}
Before feeding the aberrated images to the restoration network, we deconvolve the image first with multiple Wiener filters~\cite{wiener}. Wiener filter is a linear filter that has been used to reduce noise in signals, in the context of image processing and restoration~\cite{dong2020deep,dong2021dwdn, zhang2023infwide, zhou2021image,pronina2020microscopy,hiller1990iterative}. It can balance noise suppression and detail preservation by exploiting the signal-to-noise ratio (SNR), which makes it particularly effective in image restoration tasks. Given a PSF, the Wiener filter in the frequency domain, $W$, is defined as follows,
\begin{equation}
\label{eq:wiener_filter}
%W(u,v) = \frac{H^*(u,v)}{{|H(u,v)|}^2 + K},
W[u,v] = \frac{H^*[u,v]}{{|H[u,v]|}^2 + K},
\end{equation}
where $u$ and $v$ are horizontal and vertical spatial frequency respectively, $H$ is a fourier transformed PSF, $H^*$ is a complex conjugation of $H$, and $K$ is an inverse of SNR. Note that each color channel has a different PSF, but we do not explicitly denote the channel for brevity, e.g. $H_c$ or $W_c$.

\begin{figure}[]
\begin{center}
\includegraphics[width=1.0\linewidth]{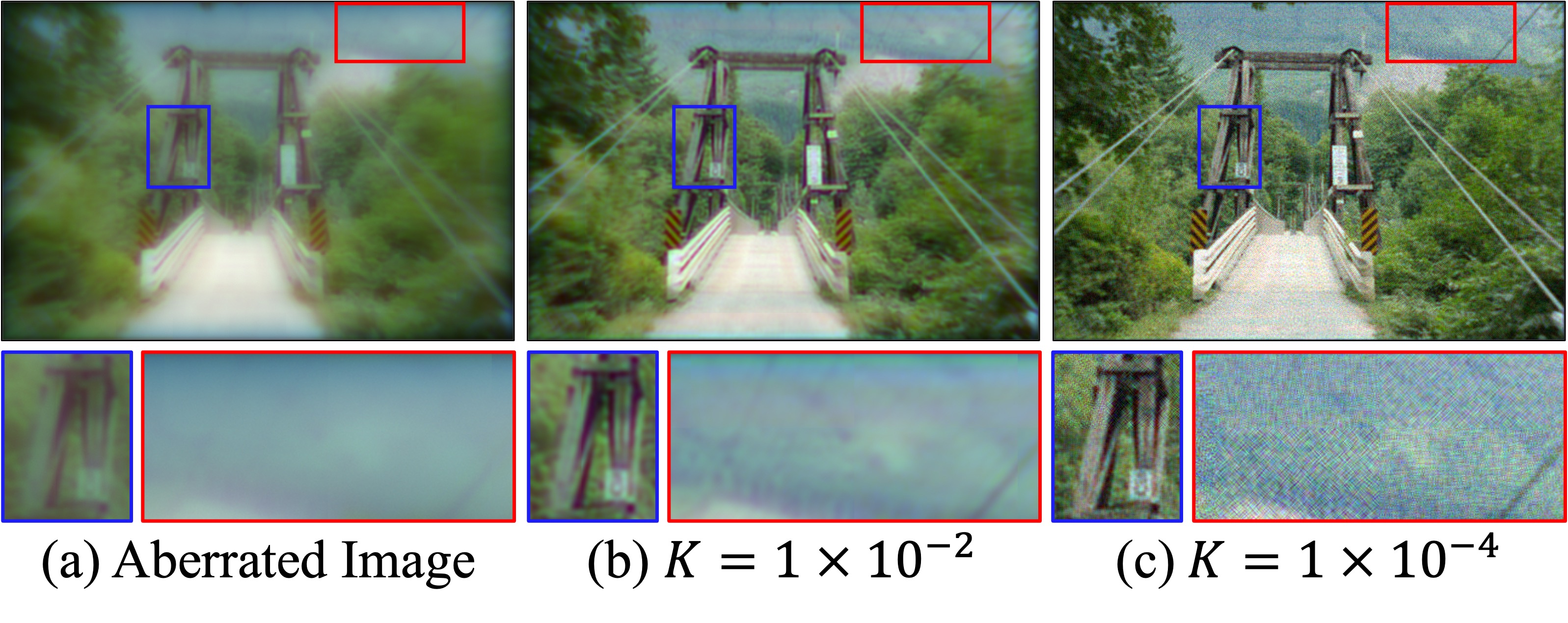}
\end{center}
\vspace{-0.8cm}
   \caption{A comparison of the Wiener deconvolved images with different noise-penalization terms $K$. High $K$ produces smooth representation (b), while smaller $K$ results in a more textured but noisy representation (c).}
\vspace{-0.3cm}
\label{fig:K_comparison}
\end{figure}
\begin{figure}[]
\begin{center}
\includegraphics[width=1.0\linewidth]{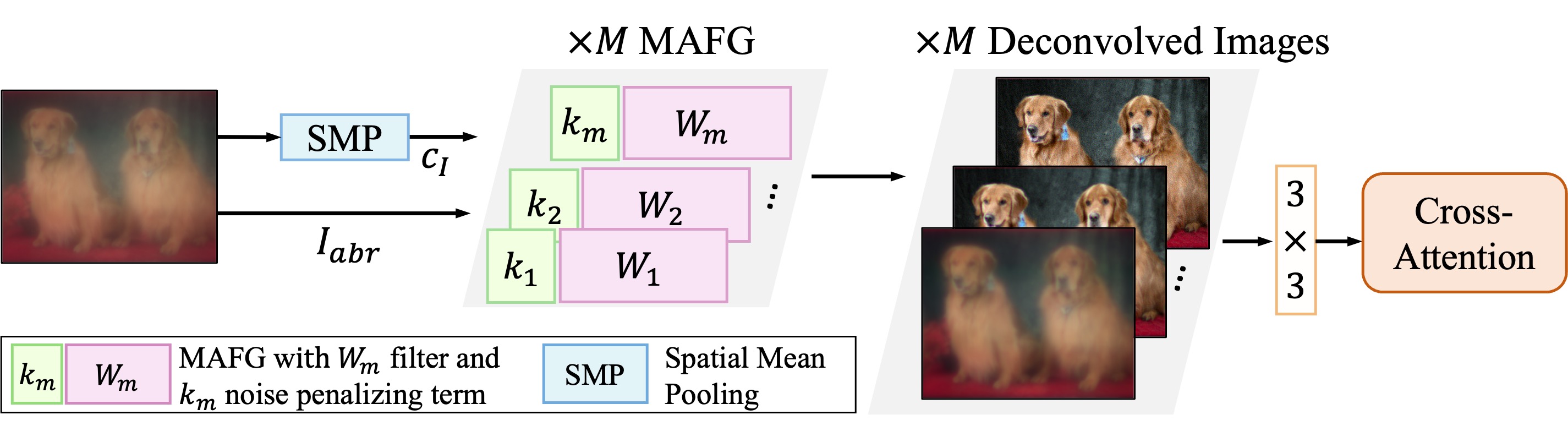}
\end{center}
\vspace{-0.55cm}
   \caption{The process of Multiple Adaptive Filters Guidance (MAFG). It produces $M$ different representations from the input images using $M$ different Wiener filters and applies cross-attention to reweight them.}
\vspace{-0.5cm}
\label{fig:MAFG}
\end{figure}

As shown in ~\cref{fig:K_comparison}, $K$ plays a pivotal role in penalizing the noise; a greater $K$ leads to a smoother representation, whereas a smaller $K$ can preserve more high-frequency details, albeit with the increased noise.
Inspired by this, we propose to use multiple Wiener filters to guide aberration correction with several distinct representations.
It is not feasible to obtain an optimal Wiener filter with accurate SNR as noise distribution is unknown in the real world.
% We adopt $M$ Wiener filters with different $K$ in a wide range to deal with obscure noise distribution.
Such design can approximate the ideal filter by employing $M$ filters with different $K$ in a wide range.

Furthermore, we extend multiple Wiener filters to Multiple Adaptive Filters Guidance (MAFG), illustrated in~\cref{fig:MAFG}, which determines $K$ adaptively considering the image intensity. Image with higher intensity tends to have better signal quality, so brighter channels often experience less noise~\cite{erkmen2009signal}. Thus, we penalize noise less and capture more information for bright images by adjusting $K$ with the image intensity. Also, we treat each channel differently to avoid suppressing high-SNR details unnecessarily, since metalens exhibit wavelength-dependent aberrations where each color channel suffers differently due to the different chromatic responses. Based on this observation, we slightly modify ~\cref{eq:wiener_filter} for MAFG as follows:
\begin{equation}
\label{eq:MAFG}
%W_m(u,v) = \frac{H^*(u,v)}{{|H(u,v)|}^2 + k_m\cdot (1- c_I)},
W_m[u,v] = \frac{H^*[u,v]}{{|H[u,v]|}^2 + k_m\cdot (1- c_I)},
\end{equation}
where $m \in \{1, \dots, M \}$ is the index of $M$ different filters, $c_I \in \mathbb{R}$ is the average intensity of each color channel, and $k_m \in \mathbb{R}$ is a hyperparameter for scaling $c_I$.
The proposed MAFG is adaptive to the input images, which allows flexible responses to the various images.

We obtain $M$ different deconvolved images $\{ I_{dec}^{m} \}_{m=1}^M$ using MAFG with the following equation:
\begin{equation}
\label{eq:MAFG_deconv}
 I_{dec}^{m}[x,y] = \texttt{ifft}(W_m \odot \texttt{fft}(I_{abr}))[x,y],
\end{equation}
where $\texttt{fft}$ and $\texttt{ifft}$ are Fast Fourier Transform (FFT) and the inverse FFT~\cite{fft}, respectively. `$\odot$' is the Hadamard product, $x$ and $y$ are pixel coordinates. 

Then, they are processed by a cross-attention block, where the query is $I^{k_{med}}_{dec}$, the deconvolved image with a median of $k_m$, which is a compromise between noise suppression and detail preservation. Cross-attention can reweight the features, which it can attend more to the filter outputs with high $k_m$ (severe aberration, red boxes in~\cref{fig:K_comparison}) or it can favor the filter outputs with low $k_m$ (mildly aberrated parts, blue boxes in~\cref{fig:K_comparison}), resulting in spatially adaptive correction. This is especially important for metalens images, where aberration occurs non-uniformly. The reweighted feature is then fed to the restoration network.

\subsection{Spatial and Transposed Self-Attention Fusion}
\label{sec:method_STAF}

\begin{figure}[t]
\begin{center}
\includegraphics[width=1.0\linewidth]{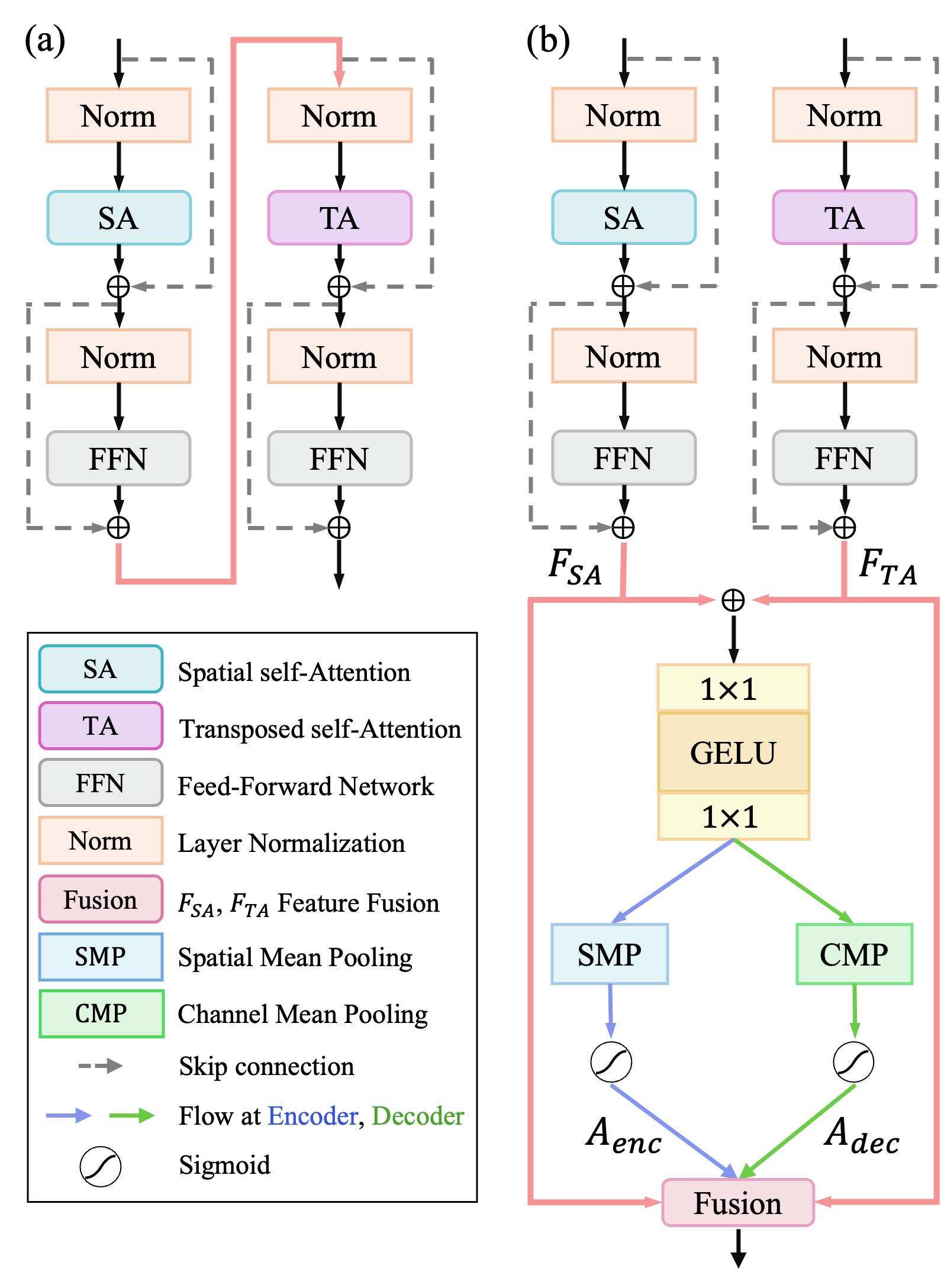}
\end{center}
\vspace{-0.5cm}
   \caption{Comparison on applying SA and TA. (a). Previous works that apply SA and TA alternatively. (b). Proposed Spatial and Transposed self-Attention Fusion (STAF). It implements SA and TA in parallel and fuses features with different weights in the encoder and decoder.}
\label{fig:STAF}
\vspace{-0.5cm}
\end{figure}
As aforementioned, a variety of previous studies adopted spatial self-attention (SA)~\cite{swinir,hat}, transposed self-attention (TA)~\cite{zamir2022restormer}, or both~\cite{xrestormer,dat,ipt} for image restoration. Leveraging both SA and TA, by alternatively applying them as depicted in~\cref{fig:STAF} (a), can encode spatial context effectively.

Nevertheless, there remains room for improvement, as existing methods apply SA and TA uniformly across both encoder and decoder stages despite their distinct objectives. The encoder primarily focuses on capturing a global context, emphasizing the overall structure and relationships within the images, which helps identify patterns and features.
% that may be corrupted in the degraded images. 
In contrast, the decoder mainly aims to recover local details and textures essential for high-fidelity restoration.

Therefore, we propose a Spatial and Transposed self-Attention Fusion (STAF) module which implements SA and TA separately, not alternatively, and fuses SA and TA features with different weights in the encoder and decoder, as depicted in~\cref{fig:STAF} (b). 
In detail, the outcome of SA and TA (denoted as $F_{SA}$,  $F_{TA} \in \mathbb{R}^{H\times W\times C}$, respectively and $H, W, C$ are height, width, and channel of the input feature) are concatenated along channel dimension to obtain concatenated feature $F \in \mathbb{R}^{H \times W \times 2C}$.
$F$ is then fed to a small CNN, consisting of two 1$\times$1 convolution layers with GELU activation~\cite{gelu}, to mix features and halve channel dimension to $C$, and produce $A \in \mathbb{R}^{H \times W \times C}$, which is then treated differently in encoder and decoder.
At the encoder side, we employ average pooling to $A$ on spatial dimensions to efficiently reduce resolution while preserving crucial feature representations and obtain weight matrix $A_{enc} \in \mathbb{R}^{1 \times 1 \times C}$.
On the other hand, $A$ is averaged along the channel dimension to selectively emphasize the relevant feature maps and produce a weight matrix $A_{dec} \in \mathbb{R}^{H \times W \times 1}$ for the decoder.
This allows the decoder to prioritize different spatial areas.
% , for instance, regions with more aberration can receive higher attention to be corrected.
In a nutshell, $A_{enc}$ and $A_{dec}$ can be computed as follows:
\begin{equation}
\label{eq: STAF}
\begin{aligned}
    F = \texttt{Concat}(F_{SA}, F_{TA}), \\
    A = \texttt{Conv}(\texttt{GELU}(\texttt{Conv}(F))), \\
    A_{enc} = \sigma (\texttt{AVGPool}_{H,W}(A)), \\
    A_{dec} = \sigma (\texttt{AVGPool}_{C}(A)), \\
\end{aligned}
\end{equation}
where $\sigma$ is the sigmoid activation, and $\texttt{AVGPool}_{H,W}$ and $\texttt{AVGPool}_{C}$ are average pooling along the height and width dimension and channel dimension, respectively.
$A_{enc}$ and $A_{dec}$ are then used to compute the final feature of each block where $F_{SA}$ and $F_{TA}$ are mixed properly for the encoder and decoder with the equations:
\begin{equation}
\label{eq: STAF_sum}
\begin{aligned}
    F_{enc} = F_{SA} \odot A_{enc} + F_{TA} \odot (1-A_{enc}), \\
    F_{dec} = F_{SA} \odot A_{dec} + F_{TA} \odot (1-A_{dec}), \\
\end{aligned}
\end{equation}
where $\odot$ denotes the Hadamard product, and the index for each layer is omitted for brevity. 

To further compensate for fine detail losses due to the harsh aberration, we enforce the queries of both attention modules in the decoder to focus more on high-frequency information, inspired by ESRT~\cite{lu2022transformersingleimagesuperresolution} that estimates high-frequency information of the feature using average pooling. In particular, we adopt this enhanced query $Q_{dec}^b$ in every other block of each decoder layer, as follows:
\begin{equation}
\label{eq:query_enhancement}
Q_{dec}^b =
\begin{cases}
    Q_{dec}'^b, & \text{if } b = 2n \\
    Q_{dec}'^b - \texttt{AVGPool}(Q_{dec}'^b), & \text{if } b = 2n+1, \\
\end{cases}
\end{equation}
where $b$ is a block index, $n \in \mathbb{Z}$, $Q_{dec}'^b$ is an original query from the previous block, and the index of the layer is again omitted for brevity. This enhanced query is used in the SA and TA blocks of the decoder to compute $F_{dec}$. Finally, the residual obtained after forwarding all STAF blocks is added to the $I_{dec}^{k_{med}}$, to get the corrected image $I_{ac}$.

\begin{figure}
\begin{center}
\includegraphics[width=1.0\linewidth]{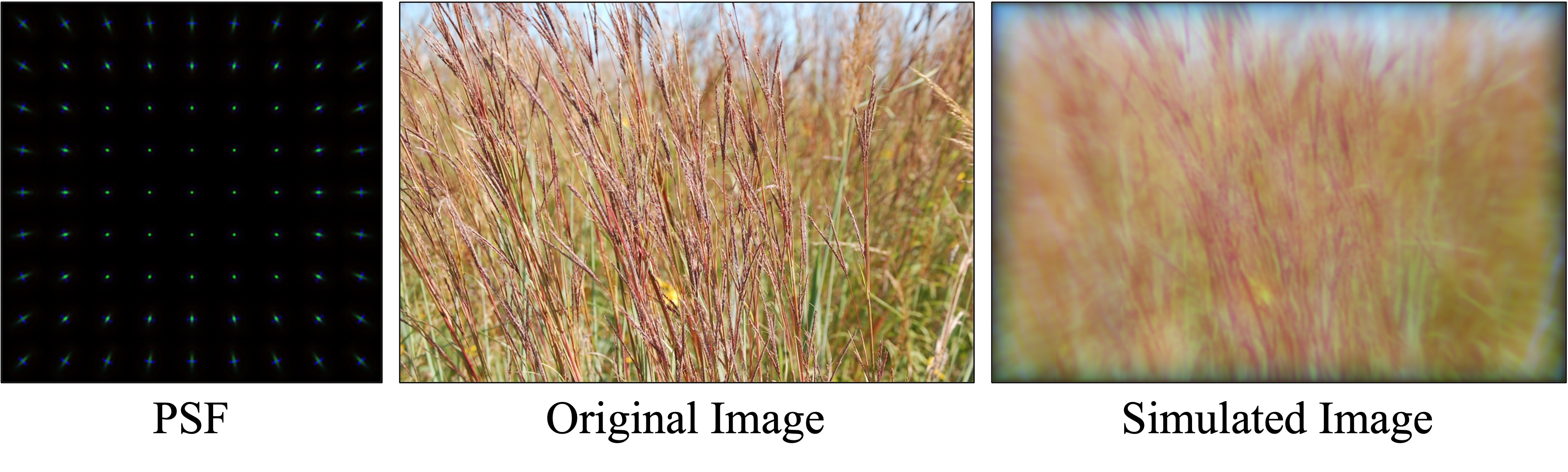}
\end{center}
\vspace{-0.6cm}
   \caption{Generated point spread functions (PSFs) and an image with spatially varying aberration derived from them.}
\vspace{-0.4cm}
\label{fig:simul_psf}
\end{figure}
\begin{figure*}[]
\begin{center}
\includegraphics[width=1.0\linewidth]{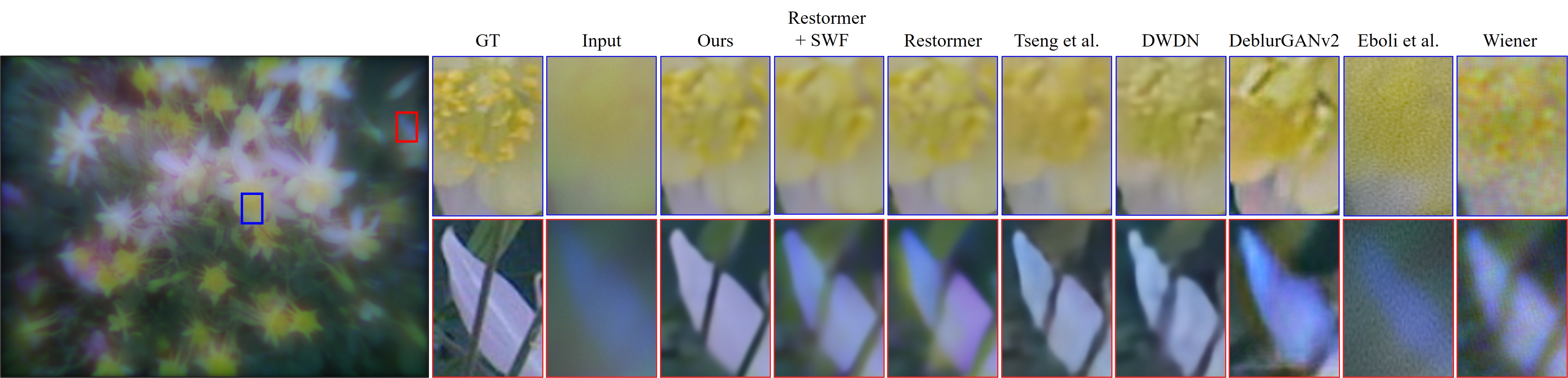}
\end{center}
\vspace{-0.6cm}
   \caption{Qualitative results on the Open Image V7~\cite{openv4,openv7} dataset with spatially varying aberration.}
\vspace{-0.4cm}
\label{fig:quali_open_varying}
\end{figure*}
\begin{figure*}[]
\begin{center}
\includegraphics[width=1.0\linewidth]{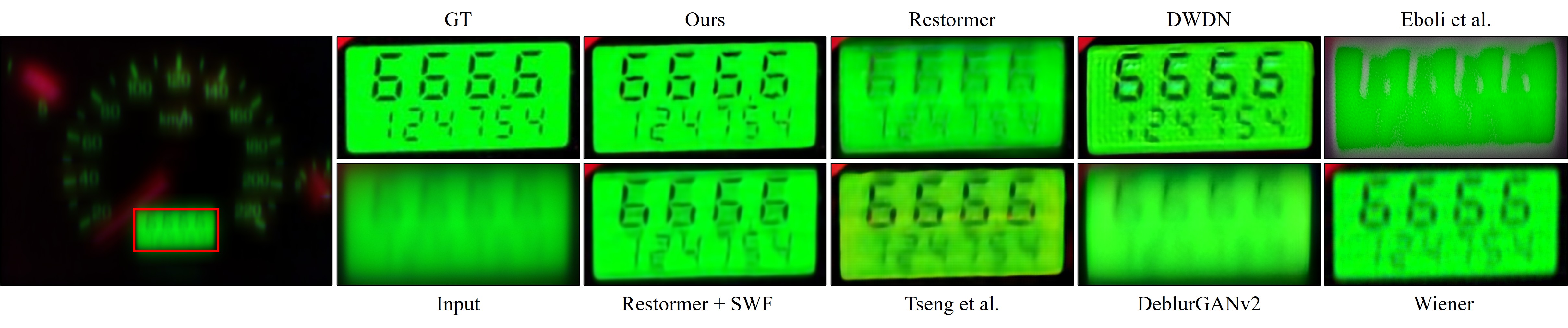}
\end{center}
\vspace{-0.8cm}
   \caption{Qualitative results on the Open Image V7~\cite{openv4,openv7} dataset with the uniform aberration at field angle 20\textdegree.}
% \vspace{-0.3cm}
\label{fig:quali_open_invarying}
\end{figure*}

\begin{figure*}[]
\begin{center}
\includegraphics[width=1.0\linewidth]{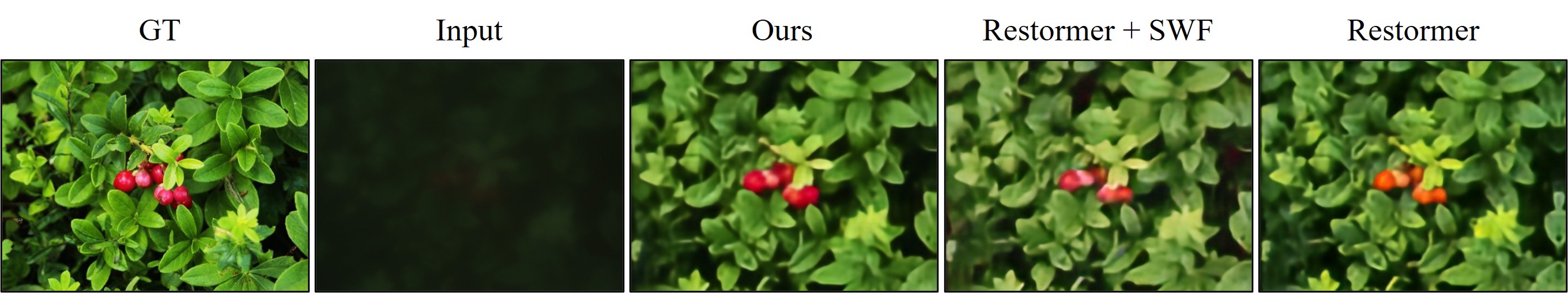}
\end{center}
\vspace{-0.6cm}
   \caption{Qualitative results on the images captured with our fabricated metalens.}
\vspace{-0.4cm}
\label{fig:quali_real}
\end{figure*}

\begin{figure*}
\begin{center}
\includegraphics[width=1.0\linewidth]{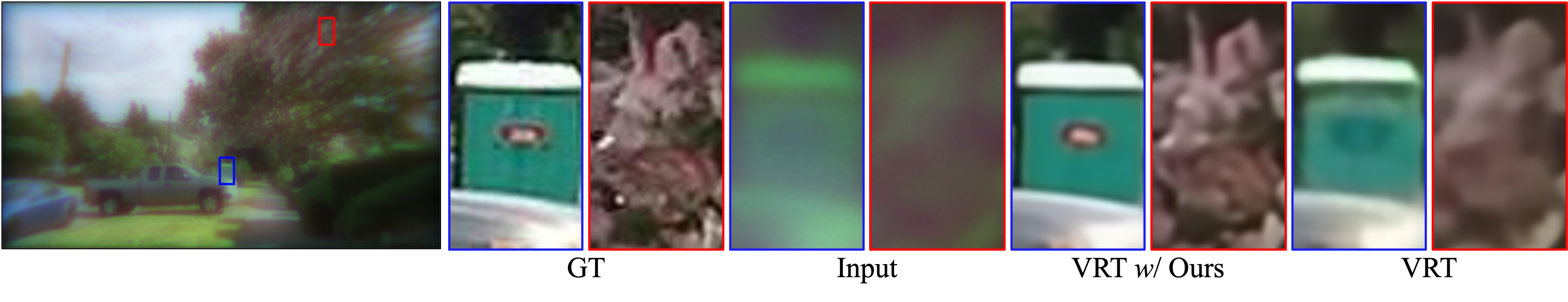}
\end{center}
\vspace{-0.8cm}
   \caption{Qualitative results on the aberrated DVD~\cite{dvd} dataset.}
\vspace{-0.5cm}
\label{fig:quali_video}
\end{figure*}

\label{sec:exp_synthetic}
\begin{table*}[t!]
    \centering
    \resizebox{\textwidth}{!}{
    \begin{tabular}{c|ccc|ccc|ccc}
    \hline
    & \multicolumn{3}{c|}{Non-uniform aberration} & \multicolumn{3}{c|}{Uniform aberration at $\theta = 0$\textdegree} & \multicolumn{3}{|c}{Uniform aberration at $\theta = 20$\textdegree} \\
    Method & PSNR$\uparrow$ & SSIM$\uparrow$ & LPIPS$\downarrow$ & PSNR$\uparrow$ & SSIM$\uparrow$ & LPIPS$\downarrow$ & PSNR$\uparrow$ & SSIM$\uparrow$ & LPIPS$\downarrow$ \\
    \hline
    \hline
    Wiener deconvolution~\cite{wiener} & 25.54 & 0.5743 & 0.5228 & 27.06 & 0.6561 & 0.4458 & 26.05 & 0.6332 & 0.5030   \\
    \citet{eboli2022fast} & 15.19 & 0.3784 & 0.8270 & 20.73 & 0.5359 & 0.6488 & 18.55 & 0.4480 & 0.7973  \\
    DeblurGANv2~\cite{kupyn2019deblurgan} & 24.08 & 0.6863 & 0.3233 & 25.27 & 0.7338 & 0.3032 & 21.79 & 0.6048 & 0.4810 \\
    DWDN~\cite{dong2021dwdn} & 26.40 & 0.7656 & 0.2854 & 29.38 & 0.8318 & 0.2459 & 25.68 & 0.7282 & 0.3267  \\
    \citet{tseng2021neural} & 28.66 & 0.8045 & 0.2949 & 29.88 & 0.8373 & 0.2507 & 28.09 & 0.7841 & 0.3117  \\
    Restormer~\cite{zamir2022restormer} & 27.84 & 0.8091 & 0.2753 & 31.09 & 0.8827 & 0.1804 & 28.53 & 0.8020 & 0.2871 \\
    Restormer~\cite{zamir2022restormer} + SWF & 29.87 & 0.8289 & 0.2812 & 33.58 & 0.8804 & 0.2003 & 28.95 & 0.7908 & 0.3262\\
    Ours & \textbf{34.29} & \textbf{0.8760} & \textbf{0.2052} & \textbf{35.06} & \textbf{0.8961} & \textbf{0.1763} & \textbf{32.32} & \textbf{0.8409} & \textbf{0.2542}   \\
    \hline
    \end{tabular}
    }
    \vspace{-0.3cm}
    \caption{Quantitative results Open Image V7~\cite{openv4,openv7} dataset with spatially varying and invariant aberration, where $\theta$ denotes a field angle.}
    \label{tab:quant_result_open}
    \vspace{-0.4cm}
\end{table*}
\section{Experiments}
\label{sec:experiments}
\subsection{Experimental Settings}
We divided FOV into $9 \times 9$ grid and generated 81 PSFs using metalens phase profile from~\citet{tseng2021neural} to represent spatially varying aberration of metalens imaging as shown in~\cref{fig:simul_psf}, following~\cref{eq:imaging}.
We set the number of filters $M=13$ and selected 13 logarithmically spaced values in the range $[1e-5,1e-2]$ for noise penalizing scalers $k_m$.
% using a base 10 logarithmic scale for noise penalizing scalers $k_m$. 

We conducted experiments across various tasks: image aberration correction (\cref{sec:exp_synthetic}), restoring images captured with the fabricated metalens (\cref{sec:exp_real}), video aberration correction (\cref{sec:exp_video}), and 3D reconstruction with aberrated images (\cref{sec:exp_nvs}).
Analysis on the method, detailed experimental settings, and more experimental results are described in the supplementary material.

\subsection{Image Aberration Correction}
We adopted  Open Image V7~\cite{openv4,openv7} dataset with spatially varying aberration for image aberration correction task, and sampled 120K and 1K images for training and validation, respectively. We constructed two different types of test sets, one with spatially varying aberrations from 81 different PSFs, and the other with spatially invariant aberration, where we selected 5 PSFs at different field angles (0\textdegree, 5\textdegree, 10\textdegree, 15\textdegree, and 20\textdegree), each corresponding to a distinct uniform aberration.
% and simulated 5 different uniform aberrations using each PSF.

\cref{tab:quant_result_open}, \cref{fig:quali_open_varying} and \cref{fig:quali_open_invarying} provide the results on the spatially varying and invariant aberration test sets\footnote{The codes and pre-trained weights for \citet{chakravarthula2023thin} were not publicly available, so we could only compare with \citet{tseng2021neural}.}, where \textit{Restormer+SWF} is a \textit{non-blindly} trained Restormer using a single Wiener filter. 
As shown in input images in \cref{fig:quali_open_invarying}, an image degraded by the PSF at a large field angle suffers more from aberration, which corresponds to the outer parts of the image with spatially varying aberration in~\cref{fig:quali_open_varying}. 

\noindent\textbf{Spatially Varying Aberration} \citet{tseng2021neural} leverages the PSF to address aberrations but fails to preserve fine details and contrast, highlighting the limitations of CNNs for metalens imaging. 
Restormer~\cite{zamir2022restormer} demonstrates the capability of ViTs in metalens imaging by restoring the mildly aberrated region (blue box in~\cref{fig:quali_open_varying})  without PSF guidance. However, it fails to correct pronounced aberrations in off-axis areas (red box in~\cref{fig:quali_open_varying}). 
% Non-blindly training Restormer produced better results, yet employing only one filter struggles with recovering fine details. 
Non-blindly training Restormer with a single Wiener filter still struggles with recovering fine details. 
In contrast, our approach effectively corrects aberrations across both on- and off-axis regions, achieving state-of-the-art results with a noticeable margin.

\noindent\textbf{Spatially Invariant Aberration} \citet{tseng2021neural}  models spatially varying aberration by utilizing multiple patches with different degrees of aberration. However, since each patch assumes a single type of aberration (i.e., uniform aberration), seam-like artifacts may emerge when applied to the image where diverse aberrations coexist within a single frame, which is common in real scenarios. To ensure a fair comparison, we further conducted experiments under uniform aberration conditions. As demonstrated in \cref{tab:quant_result_open} and \cref{fig:quali_open_invarying}, our method effectively models spatially varying aberrations, achieving robust performance even in cases where severe aberrations (at field angle 20\textdegree) appear uniformly, outperforming \citet{tseng2021neural}.

\begin{table}[]
    \centering
    \begin{tabular}{c|c|c|c}
    \hline
    Method & PSNR$\uparrow$ & SSIM$\uparrow$ & LPIPS$\downarrow$ \\
    \hline
    \hline
    VRT~\cite{vrt} & 23.23 & 0.6906 & 0.3921   \\
    VRT \textit{w/} Ours & \textbf{28.89} & \textbf{0.8602} & \textbf{0.2102} \\
    \hline
    \end{tabular}
    \vspace{-0.3cm}
    \caption{Quantitative results on the aberrated DVD dataset~\cite{dvd}. }
    \vspace{-0.4cm}
    \label{tab:quant_video}
\end{table}

\subsection{Images from the Fabricated Metalens}
\label{sec:exp_real}
We fabricated a metalens with the PSF from \citet{tseng2021neural} as shown in \cref{fig:quali_real} (Left) and captured images to construct a dataset. It consists of 220 images, where 200 images are involved in fine-tuning the pre-trained model from~\cref{sec:exp_synthetic}, and 20 images are kept for validation. Since ViT requires a larger volume of training data, large-scale dataset construction is inevitable, which is labor-intensive. Instead, we trained a model with numerous data simulated with the same phase profile of the manufactured metalens, and then fine-tuned it with the small set of captured images, easing the cost of collecting data. We fine-tuned our model, Restormer~\cite{zamir2022restormer} and non-blind Restormer under the same experimental setting. As shown in~\cref{fig:quali_real}, both Restormer and non-blind Restormer produce images in low contrast while the proposed method yields images with better contrast, leading to more visually appealing quality.
% The fine-tuned model can correct the aberration of the images captured with the metalens, as shown in~\cref{fig:quali_real}.
% effectively narrowing down the domain gap, 

% as shown in \cref{fig:quali_real} (Right).

\subsection{Video Aberration Correction}
\label{sec:exp_video}
In this section, we assessed our method for the video aberration correction task. We employed VRT~\cite{vrt} as a baseline video restoration model and trained it on the DVD~\cite{dvd} dataset with spatially varying aberration. We trained VRT with the output of pre-trained model (VRT \textit{w/} Ours) from~\cref{sec:exp_synthetic}. \cref{tab:quant_video} and~\cref{fig:quali_video} show that VRT struggled with correcting the aberration of the input video while VRT \textit{w/} Ours could enjoy both temporal consistency and clean representations. 
% \todo{blah}

\subsection{3D Reconstruction with Aberrated Images}
\label{sec:exp_nvs}
\begin{table}[]
    \centering
    \resizebox{\linewidth}{!}{
    \begin{tabular}{c|ccc|ccc}
    \hline
    & \multicolumn{3}{c|}{LLFF~\cite{llff}} & \multicolumn{3}{|c}{Tanks\&Temples~\cite{tanks}} \\
    Method & PSNR$\uparrow$ & SSIM$\uparrow$ & LPIPS$\downarrow$ & PSNR$\uparrow$ & SSIM$\uparrow$ & LPIPS$\downarrow$\\
    \hline
    \hline
    3D-GS~\cite{3dgs} & 17.43 & 0.4083 & 0.7428 & 16.81 & 0.3875 & 0.7263 \\
    3D-GS \textit{w/} Ours & 25.29 & 0.7513 & \textbf{0.2138} & 23.17 & 0.6852 & 0.2639 \\
    3D-GS + Ours & \textbf{25.49} & \textbf{0.7548} & 0.2149 & \textbf{23.30} & \textbf{0.6905} & \textbf{0.2632} \\
    \hline
    \end{tabular}
    }
    \vspace{-0.2cm}
    \caption{Quantitative results on the LLFF~\cite{llff} and Tanks\&Temples~\cite{tanks} dataset with spatially varying aberration.}
    \vspace{-0.2cm}
    \label{tab:quant_nvs_llff_n_tnt}
\end{table}

\begin{figure}
\begin{center}
\includegraphics[width=1.0\linewidth]{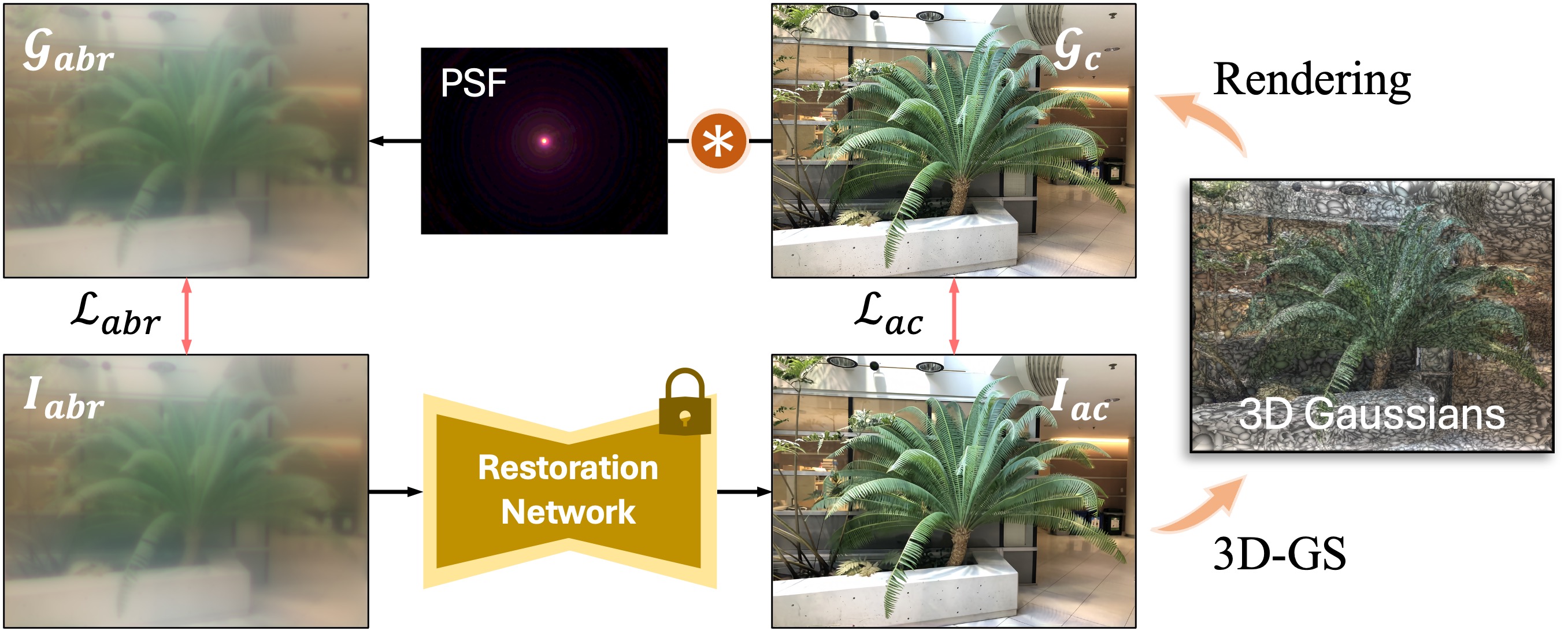}
\end{center}
\vspace{-0.5cm}
   \caption{Training pipeline for 3D-GS using the proposed ViT.}
   % $*$ denotes a convolution operation.}
\vspace{-0.4cm}
\label{fig:3dgs}
\end{figure}

\begin{figure}
\begin{center}
\includegraphics[width=1.0 \linewidth]{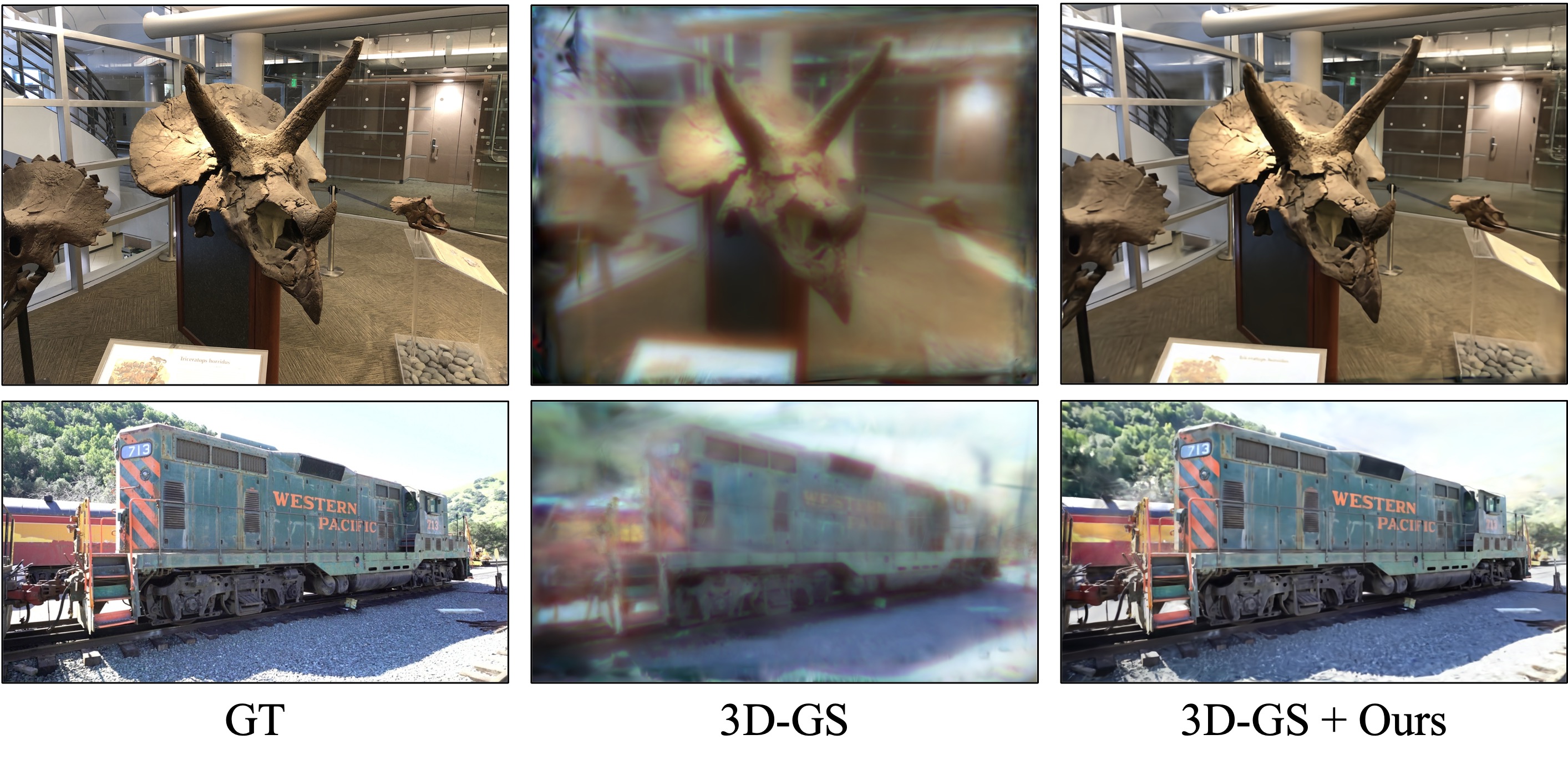}
\end{center}
\vspace{-0.7cm}
   \caption{Qualitative results on the aberrated LLFF~\cite{tanks} (Top) and Tanks\&Temples~\cite{tanks} (Bottom) dataset.}
\vspace{-0.5cm}
\label{fig:quali_nvs_llff_n_tnt}
\end{figure}

We further extended our method to clean 3D reconstruction, training 3D Gaussian Splatting (3D-GS)~\cite{3dgs} with the sets of multi-view aberrated images. We embedded our method into 3D-GS training pipeline as delineated in~\cref{fig:3dgs}. Specifically, we simulated aberration during 3D-GS training using~\cref{eq:imaging} to constrain 3D-GS learning clean representations.
The output ($I_{ac}$) of our pre-trained ViT, which is frozen during 3D-GS training, is involved in supervising the rendered image ($\mathcal{G}_c$) with a loss $\mathcal{L}_{ac} = \mathcal{L}_{GS}(\mathcal{G}_c, I_{ac})$ where $\mathcal{L}_{GS}$ is a reconstruction loss function defined in 3D-GS.
The aberration-simulated image $\mathcal{G}_{abr}$ is used to compute a loss $\mathcal{L}_{abr} = \mathcal{L}_{GS}(\mathcal{G}_{abr}, I_{abr})$, with the training image $I_{abr}$. We use a sum of these losses, $\mathcal{L}_{train} = \mathcal{L}_{abr} + \lambda\mathcal{L}_{ac}$, to guide 3D-GS learning clean features from the aberrated images, where $\lambda$ is a hyperparameter.

\cref{tab:quant_nvs_llff_n_tnt} and provide the results on the 3D reconstruction task, evaluated on LLFF~\cite{llff} and Tanks\&Temples~\cite{tanks} dataset with spatially variant aberration where we used only \textit{Truck} and \textit{Train} scenes in Tanks\&Temples dataset for evaluation, following the experimental setting in 3D-GS.
While 3D-GS fails to render clean images, the proposed pipeline that combines 3D-GS and our image restoration model (3D-GS + Ours) shows sound reconstruction performance, even outperforming 3D-GS optimized with the restored output $I_{ac}$ from the proposed image correction model (3D-GS \textit{w/} Ours). This is because 3D-GS + Ours can learn clean representation via $\mathcal{L}_{ac}$ and multi-view consistency with $\mathcal{L}_{abr}$ simultaneously, resulting in better reconstruction quality.
% This is because even though our pre-trained restoration network can offer aberration-corrected representations, multi-view consistency is not considered during the aberration correction process and such 3D inconsistency in the training images leads to noisy rendering results. Meanwhile, 3D-GS + Ours can learn clean representation via $\mathcal{L}_{ac}$ and meet the multi-view consistency with $\mathcal{L}_{abr}$ simultaneously, resulting in better reconstruction quality. 
\cref{fig:quali_nvs_llff_n_tnt} shows qualitative results, where the proposed method can produce clean images in novel views, while vanilla 3D-GS still suffers from severe aberration.

\section{Conclusion}
\label{sec:conclusion}
We present an aberration correcting transformer tailored for metalens imaging. We harness a Vision Transformer (ViT) that has not been fully explored in metalens imaging.
We devise Multiple Adaptive Filters Guidance (MAFG) that can better model spatially varying aberration of metalens images and Spatial and Transposed self-Attention Fusion (STAF) which exploits features from SA and TA considering the roles of the encoder and decoder. The comprehensive experiments show the proposed method achieving state-of-the-art performance across diverse tasks.
% , outperforming previous arts by a noticeable margin. 
We also fabricate a metalens and effectively restore the captured images, proving the practicality of the proposed method.
{
    \small
    \bibliographystyle{ieeenat_fullname}
    \bibliography{main}
}

\clearpage
\appendix
\renewcommand\thesection{\Alph{section}}
\setcounter{section}{0}

\twocolumn[
\begin{center}
\Large{\bf{Aberration Correcting Vision Transformers for High-Fidelity Metalens Imaging: \\ Supplementary Materials}}\par\vspace{3ex}
\end{center}]

\section{Metalens Fabrication}
\subsection{Fabrication Details}

A metalens with a diameter of 500 $\mu$m and a focal length of 1 mm was designed based on the optimization of a polynomial phase equation \cite{tseng2021neural}. The SiN meta-atom library was generated using rigorous coupled-wave analysis (RCWA) simulations for circular pillars with a height of 750 nm. The widths of the selected meta-atoms ranged from 100 to 300 nm, with a lattice period of 350 nm.

\begin{figure}[]
\begin{center}
\includegraphics[width=1.0\linewidth]{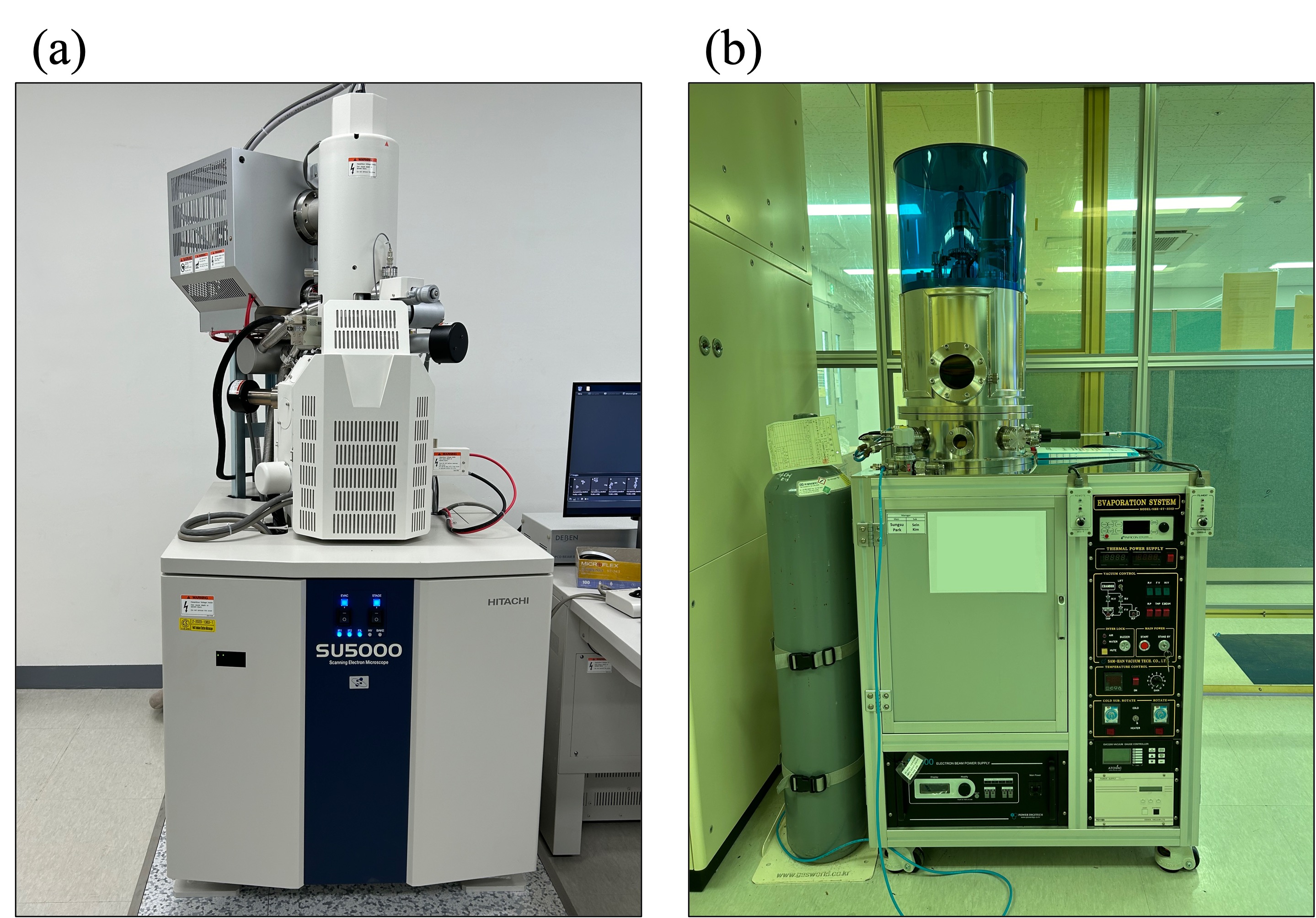}
\end{center}
\vspace{-0.5cm}
   \caption{Equipments used for metalens fabrication. (a). Electron beam lithography system for high precision patterning of the metalens. (b). Electron beam evaporator to deposit a Cr layer as a hard mask during the fabrication process.}
\vspace{-0.2cm}
\label{fig:fab}
\end{figure}

A 750 nm thick SiN layer was deposited onto a SiO$_2$ substrate using plasma-enhanced chemical vapor deposition (PECVD; Oxford, PlasmaPro 100 Cobra) to fabricate the designed metalens. A 200 nm thick positive photoresist layer (AR-P 6200.09, Allresist) was spin-coated at 4000 RPM. The pattern of circular nano-pillar meta-atoms was then transferred onto the positive photoresist using electron beam lithography, as shown in \cref{fig:fab} (a), with a dose of 3.75 C/m$^2$. To prevent charging, 100 $\mu$L of ESPACER (RESONAC, 300Z) was spin-coated at 2000 RPM for 30 s.

\begin{figure}[]
\begin{center}
\includegraphics[width=1.0\linewidth]{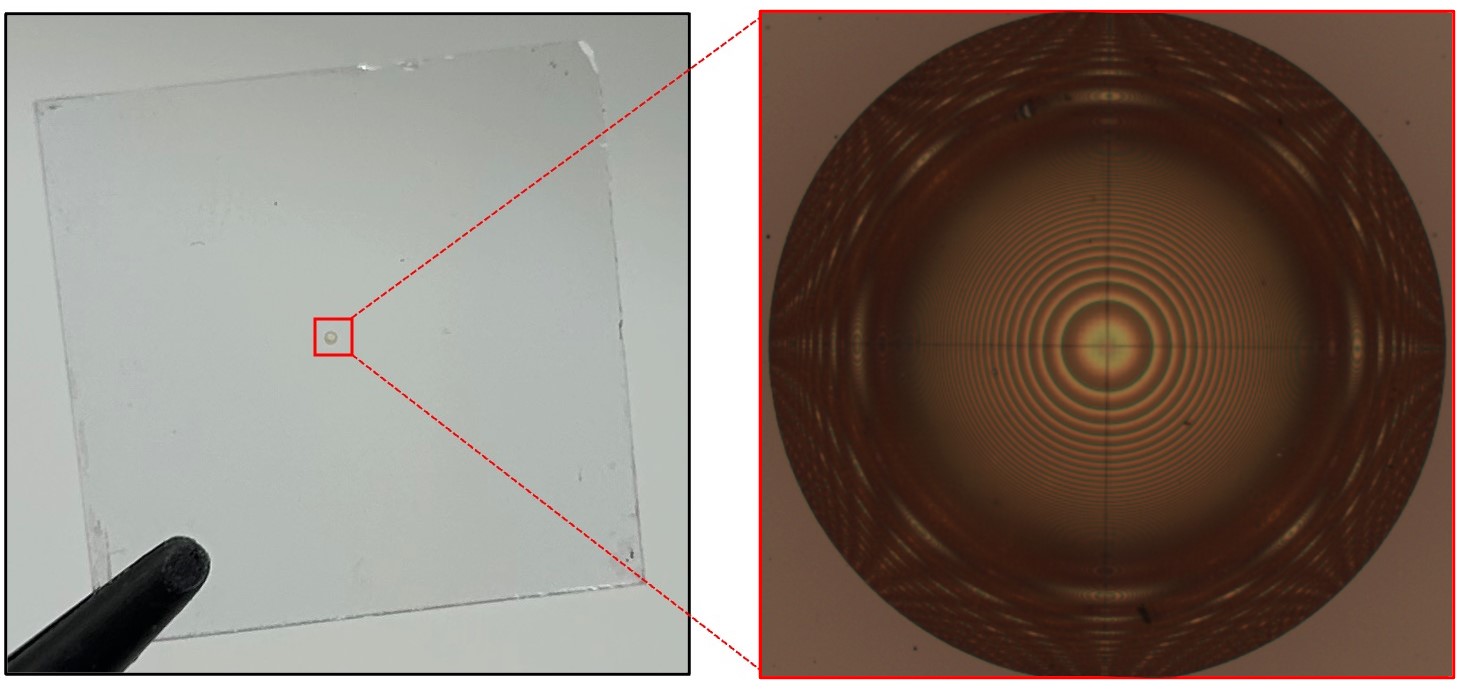}
\end{center}
\vspace{-0.5cm}
   \caption{The fabricated metalens.}
\vspace{-0.5cm}
\label{fig:metalens}
\end{figure}

The exposed resist was developed in a 1:3 solution of methyl isobutyl ketone (MIBK)/isopropyl alcohol (IPA) for 11 min. Subsequently, a 40 nm thick chromium (Cr) layer was deposited as a hard mask using an electron beam evaporator (\cref{fig:fab} (b)). The unexposed photoresist was removed through a lift-off process in acetone at room temperature for 1 hour, leaving the Cr hard mask intact. Patterning was finalized using inductively coupled plasma (ICP) etching (STS, multiplex ICP) with SF$_6$ (15 sccm) and C$_4$H$_8$ (40 sccm) gases for 10 min. Finally, the Cr hard mask was removed using a chromium etchant (TRANSENE, CE-905N) for 5 min. The fabricated metalens is illustrated in \cref{fig:metalens}.

\begin{figure*}[!h]
\begin{center}
\includegraphics[width=1.0\linewidth]{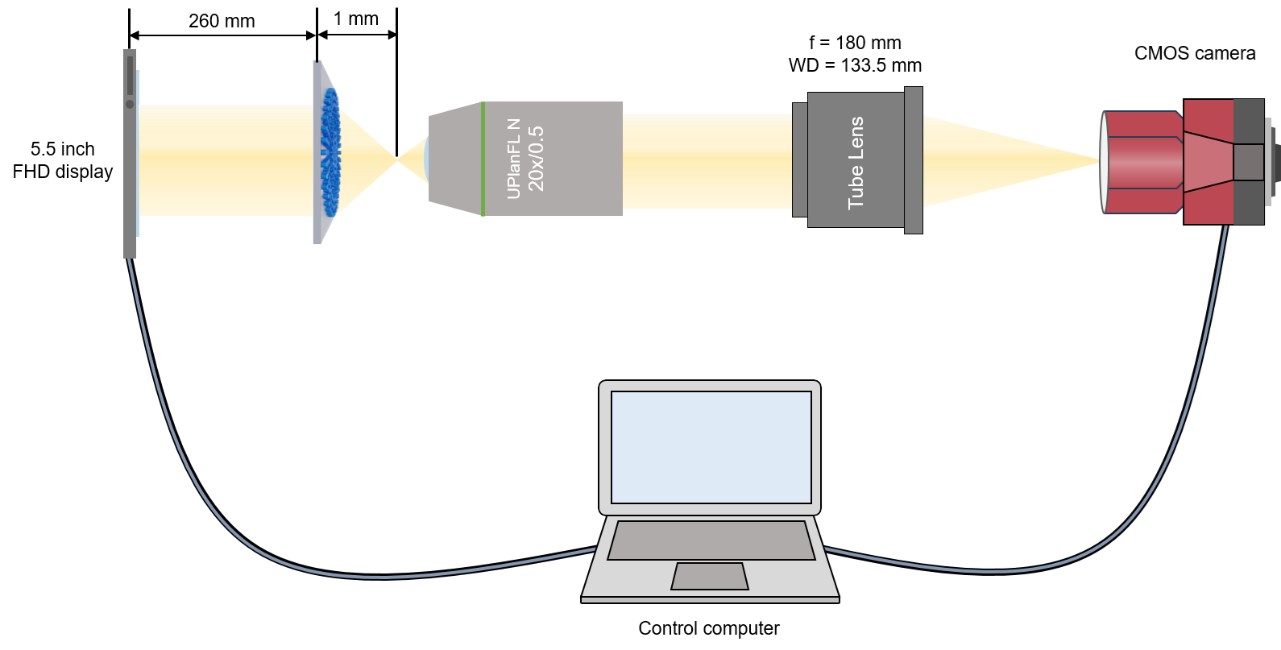}
\end{center}
\vspace{-0.2cm}
   \caption{Experimental display imaging setup. The same setup was used to capture RGB PSFs.}
% \vspace{0.3cm}
\label{fig:imaging_setup}
\end{figure*}

\subsection{Optical Setup for Imaging}
An image capture setup is illustrated in~\cref{fig:imaging_setup}.
An optical microscope system was set up to obtain images through the metalens. The images displayed on a 5.5-inch FHD display (FeelWorld, LUT5) were captured using a CMOS camera (Allied Vision, Alvium 1800 U-235c) coupled with a magnification system consisting of a 20x objective lens (Olympus, UPlanFL N 20x) with 0.5 NA and a tube lens (Thorlabs, TTL180-A). 

The metalens was positioned such that its focal plane coincided with the focal plane of the objective lens using a linear motorized stage (Thorlabs, DDS100). Camera exposure time was adjusted using a white image prior to recording to prevent saturation. 

The point spread functions (PSFs) were then acquired using the same setup with 450 nm laser (Thorlabs, CPS450), 532 nm laser (Thorlabs, CPS532), and 635 nm laser (Thorlabs, CPS635) for calibration and training of the model.

\section{Analysis and Discussion}
In this section, we provide a comprehensive analysis to demonstrate the effectiveness of the proposed methods. All models are trained with a quarter of the original batch size.

\noindent\textbf{Effect of MAFG.} To validate the effectiveness of MAFG, we conducted ablation studies on each module of MAFG as shown in~\cref{tab:abl_MAFG}, which compares MAFG and other two baselines. The first baseline (\textit{MAFG w/o CA}) does not apply cross-attention to the deconvolved images but simply concatenates them and feeds the concatenated features to the restoration network, which drops PSNR by 0.28 dB. We assume this is because cross-attention can reweight the deconvolved representations considering the degree of aberrations, proving it is proper to handle metalens images where aberration varies across the image. The other baseline (\textit{MAFG w/o CI}) does not consider channel intensity when designing Wiener filters. Noise is signal-dependent (\eg bright regions have more noise variance) and requires different noise regularization on different intensities, but this approach cannot reflect such property. It undermines perceptual quality, degrading LPIPS by 0.0038. We believe that a more sophisticated approach to incorporating image information into the filters can further improve the restoration quality. 
Lastly, \cref{tab:abl_num_filters} compares the number of filters $M$ involved in MAFG. The more filters we employed, the aberration correction performance was improved as it could enrich the input of the restoration network more and reweight the features more adaptively. Thus, we use 13 different filters considering the memory. 

\begin{table}[t!]
    \centering
    % \resizebox{\linewidth}{!}{
        \begin{tabular}{c|c|c|c}
        \hline
        Method & PSNR$\uparrow$ & SSIM$\uparrow$ & LPIPS$\downarrow$\\
        \hline
        \hline
        MAFG w/o CA & 33.59 & 0.8697 & 0.2044 \\
        MAFG w/o CI & 33.82 & 0.8728 & 0.2074  \\
        MAFG & \textbf{33.87} & \textbf{0.8757} & \textbf{0.2036}  \\
        \hline
        \end{tabular}
        % }
    % \vspace{-0.2cm}
    \caption{Ablation study on the MAFG.}
    \label{tab:abl_MAFG}
\end{table}
\begin{table}[]
        \centering
        % \resizebox{\linewidth}{!}{
        \begin{tabular}{c|c|c|c}
        \hline
        \# filters & PSNR$\uparrow$ & SSIM$\uparrow$ & LPIPS$\downarrow$\\
        \hline
        \hline
        3 & 33.60 & 0.8725 & \textbf{0.2027}  \\
        7 & 33.76 & 0.8737 & 0.2040  \\
        13 &\textbf{33.87} & \textbf{0.8757} & 0.2036 \\
        \hline
        \end{tabular}
        % }
        % \vspace{-0.2cm}
        \caption{Ablation study on the number of filters of MAFG.}
        \label{tab:abl_num_filters}
\end{table}

\noindent\textbf{Effect of STAF.} \cref{tab:abl_STAF} provides experimental results to demonstrate the competence of STAF. Employing transposed self-attention (TA) alone, similar to Restormer~\cite{zamir2022restormer} often lacks encoding spatial context, leading to the worst performance among the baselines (first row in~\cref{tab:abl_STAF}). In the case of harnessing both spatial self-attention (SA) and TA, the way of fusing two different features also affects the restoration quality. Sequentially applying SA and TA (second row) where several previous works~\cite{xrestormer,dat,ipt} proposed, still fall short of performance as SA and TA are identically implemented in both encoder and decoder. Meanwhile, the proposed STAF (fourth row) can effectively aggregate the outputs from two different modules considering the roles of encoder and decoder. It further facilitates restoration by enforcing queries of attention modules in decoder blocks focusing more on the fine details and acheieves 0.18 dB improvement in PSNR compared to using original queries (third row), resulting in superior restoration quality.

\begin{table}[]
        \centering\resizebox{\linewidth}{!}{
        \begin{tabular}{c|c|c|ccc}
        \hline
        Modules & Aggregation & QE & PSNR$\uparrow$ & SSIM$\uparrow$ & LPIPS$\downarrow$ \\
        \hline
        \hline
        TA     & -            & \checkmark & 33.07 & 0.8671 & 0.2062   \\
        SA, TA & Sequential   & \checkmark & 33.63 & 0.8690 & 0.2165 \\
        % SA, TA & Sum          & \checkmark & 33.75 & 0.8700 & \textbf{0.2032} \\
        SA, TA & Weighted sum &            & 33.69 & 0.8714 & 0.2051 \\
        SA, TA & Weighted sum & \checkmark & \textbf{33.87} & \textbf{0.8757} & \textbf{0.2036} \\
        \hline
        \end{tabular}}
    % \vspace{-0.2cm}
    \caption{Ablation study on the STAF.}
    \label{tab:abl_STAF}
\end{table}

\section{Implementation Details}
We trained our model for 100K iterations with a batch size of 8 on 4 RTX A6000 GPUs.  We adopted Overlapping Cross Attention (OCA), which partitions query matrix with non-overlapping windows and key and value matrices with overlapping windows, from HAT~\cite{hat} for our SA module, following X-Restormer~\cite{xrestormer}.
We employed the Gated-Deconv Feed-Forward network from Restormer~\cite{zamir2022restormer} for our feed-forward network (FFN). Please refer to Restormer and X-Restormer for more details.

For the cross-attention module in the MAFG, we projected the deconolved images to 32-dimensional feature space. The cross-attention uses overlapping windows similar to the SA modules in STAF and adopts two attention heads. 
In the case of image dataset construction, we resized each image such that the longer dimension (either height or width) was set to 1215 pixels, while the shorter dimension was scaled proportionally to fit the images to the sensor size while preserving the original aspect ratio. We set the sensor noise $n$ in~\cref{eq:imaging} in the main paper as a sum of Gaussian and Poisson noise. Each noise is sampled from the Gaussian distribution $\mathcal{N}(x, \sigma_g^2)$ and the Poisson distribution $\mathcal{P}(\frac{x}{\sigma_p})$, respectively, where $x$ is the image, $\sigma_g=1e-5$, and $\sigma_p=4e-5$, following~\citet{tseng2021neural}.

\section{Additional Results}
In this section, we provide additional experimental results on each task that were not delivered in the main paper due to the page limit. We also attached videos of the 3D reconstruction and video aberration correction task results.

\noindent\textbf{Evaluation Metrics} We assessed the proposed method with various evaluation metrics including Peak Signal-to-Noise Ratio (PSNR), Structural Similarity Index Measure (SSIM)~\cite{ssim}, and Learned Perceptual Image Patch Similarity (LPIPS)~\cite{lpips}. We further validated our methods with non-reference metrics, including Multi-scale Image Quality (MUSIQ)~\cite{ke2021musiq}, 
Multi-dimension Attention Network Image Quality Assessment (MANIQA)~\cite{yang2022maniqa}, Hyper-network Image Quality Assessment HyperIQA~\cite{su2020blindly} and Natural Image Quality Evaluator (NIQE)~\cite{mittal2012making} on the image datasets.

\subsection{Image Aberration Correction}
\begin{figure*}[]
\vspace{0.1cm}
\begin{center}
\includegraphics[width=1.0\linewidth]{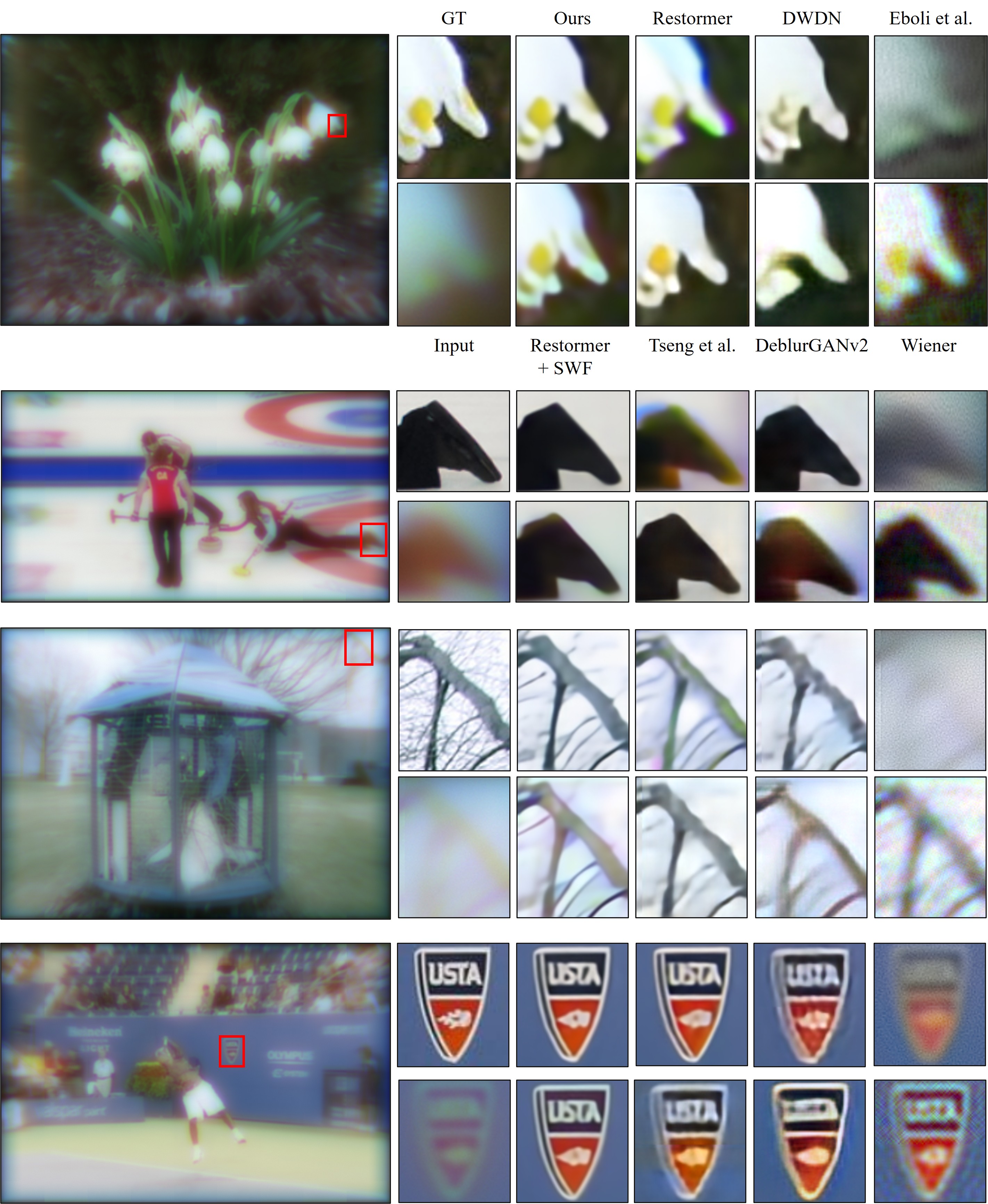}
\end{center}
\vspace{-0.3cm}
   \caption{Additional qualitative results on the Open Image V7~\cite{openv4,openv7} dataset with spatially variant aberration.}
\label{fig:varying_supp}
\end{figure*}

\cref{fig:varying_supp} shows extra qualitative results on the Open Image V7~\cite{openv4,openv7} dataset with spatially variant aberration. The proposed method can mitigate aberrations in both central and peripheral regions and restore fine details. \cref{tab:non-reference} shows the evaluations on non-uniform aberration datasets under several non-reference metrics, and the proposed method consistently outperforms diverse baseline models.

\begin{figure*}[]
\vspace{0.1cm}
\begin{center}
\includegraphics[width=1.0\linewidth]{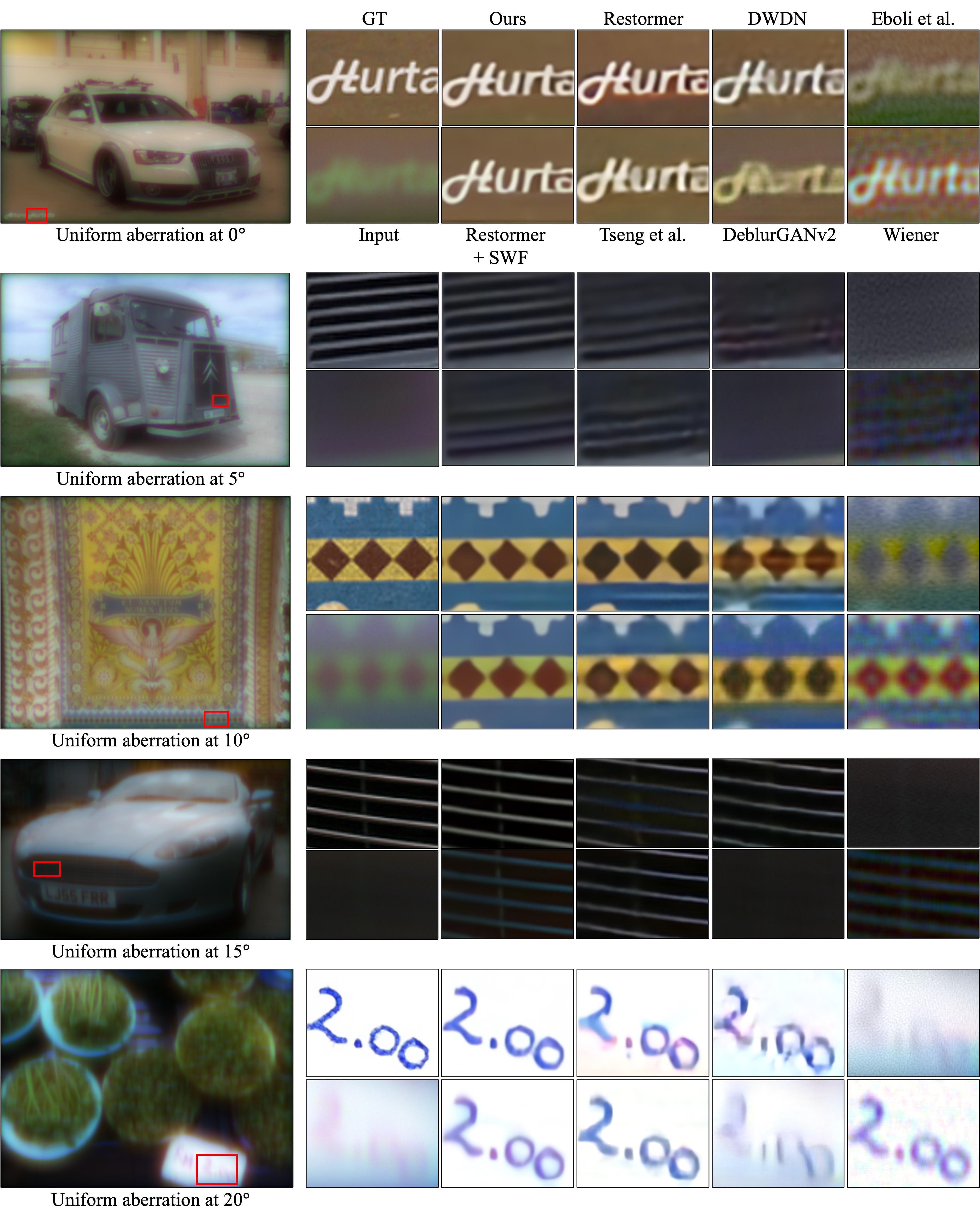}
\end{center}
\vspace{-0.6cm}
   \caption{Additional qualitative results on the Open Image V7~\cite{openv4,openv7} dataset with uniform aberration at different field angles.}
\label{fig:invarying_supp}
\end{figure*}

We also provide additional results on five different uniform aberration datasets, where we used the PSFs at 0\textdegree, 5\textdegree, 10\textdegree, and 15\textdegree, 20\textdegree~ field angle, respectively. As shown in~\cref{fig:invarying_supp}, Restormer~\cite{zamir2022restormer} shows visually satisfactory results on themildly aberrated dataset (0\textdegree~and 5\textdegree ~field angle), revealing the potential of ViT in metalens imaging. However, both Restormer and Restormer with a single Wiener filter (Restormer + SWF) struggle with restoring file details or lose image contrast when the images are harshly aberrated (15\textdegree ~and 20\textdegree ~field angle), highlighting sophisticated approach is required. Since the proposed method can model spatially variant aberration, it can also effectively correct uniform aberrations at various field angles.

\begin{table}[]
    \centering\resizebox{\linewidth}{!}{
    \begin{tabular}{c|cccc}
    \hline
    Method & MUSIQ$\uparrow$ & MANIQA$\uparrow$ & HyperIQA$\uparrow$ & NIQE$\downarrow$ \\
    \hline
    \hline
    Restormer~\cite{zamir2022restormer} & 48.028 & 0.243 &  0.340 & 7.326 \\
    Restormer~\cite{zamir2022restormer} + SWF & 48.028 & 0.239 & 0.326 & 7.295\\
    Ours & \textbf{50.938} & \textbf{0.267} & \textbf{0.360} & \textbf{6.967} \\
    \hline
    \end{tabular}}
    % \vspace{-0.3cm}
    \caption{Quantitative results on non-reference evaluation metrics.}
    \vspace{-0.3cm}
    \label{tab:non-reference}
\end{table}

\label{sec:exp_synthetic_supple}
\begin{table*}[t!]
    \centering
    \resizebox{\textwidth}{!}{
    \begin{tabular}{c|ccc|ccc|ccc}
    \hline
    & \multicolumn{3}{c|}{Uniform aberration at $\theta = 5$\textdegree} & \multicolumn{3}{c|}{Uniform aberration at $\theta = 10$\textdegree} & \multicolumn{3}{|c}{Uniform aberration at $\theta = 15$\textdegree} \\
    Method & PSNR$\uparrow$ & SSIM$\uparrow$ & LPIPS$\downarrow$ & PSNR$\uparrow$ & SSIM$\uparrow$ & LPIPS$\downarrow$ & PSNR$\uparrow$ & SSIM$\uparrow$ & LPIPS$\downarrow$ \\
    \hline
    \hline
    Wiener deconvolution~\cite{wiener} & 27.14 & 0.6653 & 0.4316 & 26.72 & 0.6455 & 0.4661 & 26.61 & 0.6479 & 0.4656\\
    \citet{eboli2022fast} & 20.75 & 0.5363 & 0.6560 & 20.36 & 0.5172 & 0.6909 & 19.40 & 0.4794 & 0.7512 \\
    DeblurGANv2~\cite{kupyn2019deblurgan} & 25.18 & 0.7286 & 0.3041 & 24.51 & 0.7025 & 0.3335 & 23.53 & 0.6599 & 0.4036  \\
    DWDN~\cite{dong2021dwdn} & 29.04 & 0.8256 & 0.2505 & 27.99 & 0.8010 & 0.2691 & 27.09 & 0.7763 & 0.2969 \\
    \citet{tseng2021neural} & 29.91 & 0.8366 & 0.2513 & 29.59 & 0.8213 & 0.2587 & 29.14 & 0.8141 & 0.2701    \\
    Restormer~\cite{zamir2022restormer} & 30.21 & 0.8654 & 0.1956 & 31.51 & 0.8671 & 0.2045 & 30.98 & 0.8526 & 0.2202  \\
    Restormer~\cite{zamir2022restormer} + SWF & 33.07 & 0.8672 & 0.2227 & 32.10 & 0.8434 & 0.2504 & 31.21 & 0.8337 & 0.2630 \\
    Ours & \textbf{34.83} & \textbf{0.8904} & \textbf{0.1795} & \textbf{34.39} & \textbf{0.8788} & \textbf{0.1979} & \textbf{34.05} & \textbf{0.8708} & \textbf{0.2070}   \\
    \hline
    \end{tabular}
    }
    \caption{Quantitative results Open Image V7~\cite{openv4,openv7} dataset with spatially varying and invariant aberration, where $\theta$ denotes a field angle.}
    \label{tab:quant_result_open_invarying_others}
    \vspace{-0.4cm}
\end{table*}

\subsection{Images from the Fabricated Metalens}
\textbf{Model Fine-tuning Details} We fine-tuned our pre-trained model, from \cref{sec:exp_synthetic} in the main paper, using images captured with the fabricated metalens. The fine-tuning dataset includes 200 images for training and 20 images for validation. However, they appeared to be dark and exhibited significant noise. To mitigate these issues, we implemented a preprocessing pipeline from \citet{shen2018improved}. Specifically, we applied Contrast-Limited Adaptive Histogram Equalization (CLAHE) \cite{109340} to enhance contrast, utilized an improved Anscombe transformation in the wavelet domain, and performed total variation-based denoising before reconstructing the images. 

We fine-tuned the pre-trained model for 5,500 iterations and the fine-tuned network could produce promising results, despite the limited dataset size.

\begin{figure*}[]
\begin{center}
\includegraphics[width=1.0\linewidth]{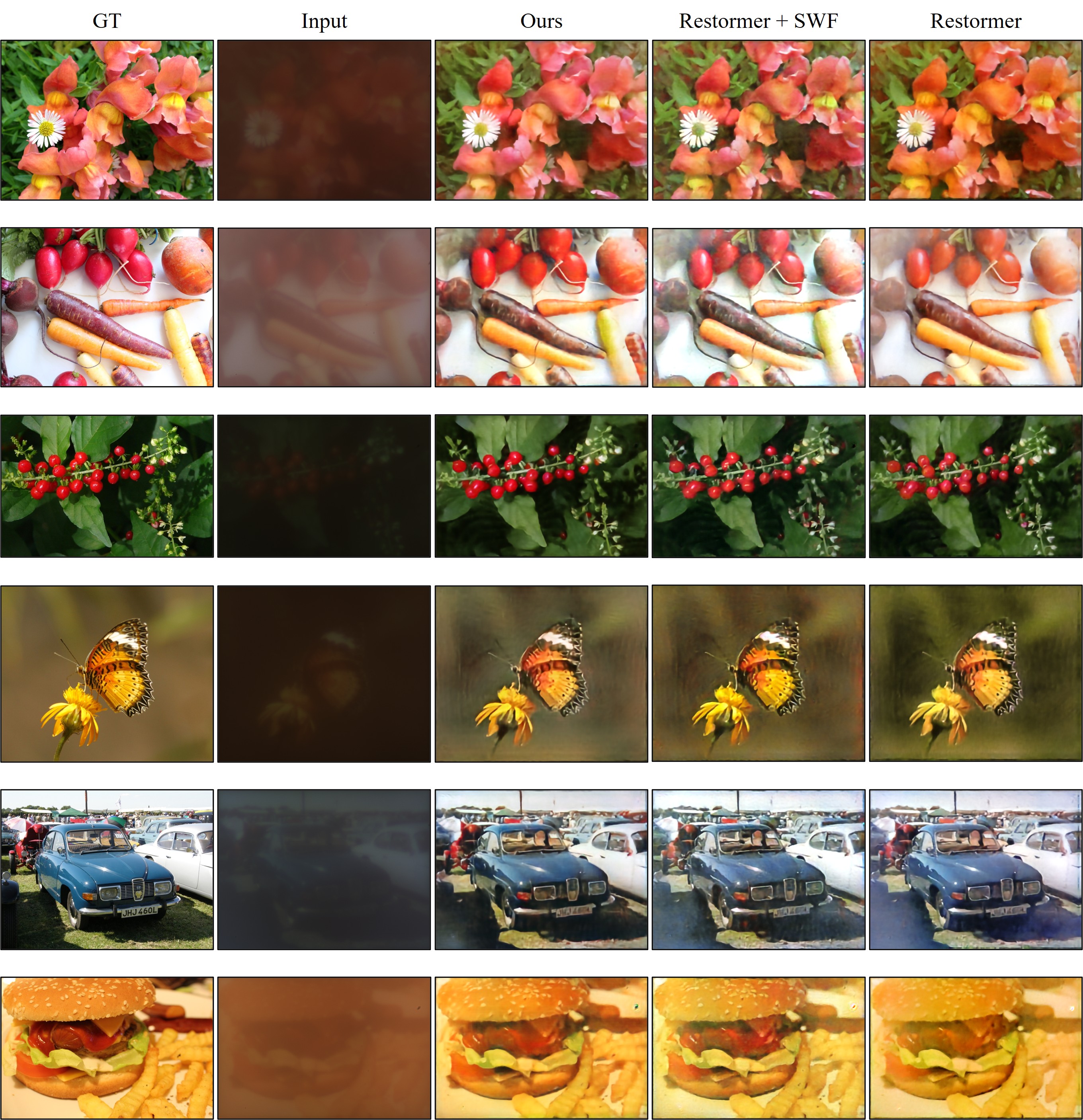}
\end{center}
% \vspace{-0.5cm}
   \caption{Additional qualitative results on the captured images with the fabricated metalens.}
% \vspace{-0.5cm}
\label{fig:real_qualitative_supp}
\end{figure*}

\noindent\textbf{Results} \cref{fig:real_qualitative_supp} presents the qualitative results of images captured with the fabricated metalens and restored ones using the proposed method and other baseline models. All models are fine-tuned under the same experimental settings. The captured images are harshly degraded, but the proposed method could effectively restore the overall color fidelity of the images with fine details, while other baseline models produce noisy and low-contrast outputs. Such outcomes can demonstrate the adaptability of the proposed method to real-world scenarios.

\subsection{Video Aberration Correction} 
\begin{figure*}[]
\begin{center}
\includegraphics[width=1.0\linewidth]{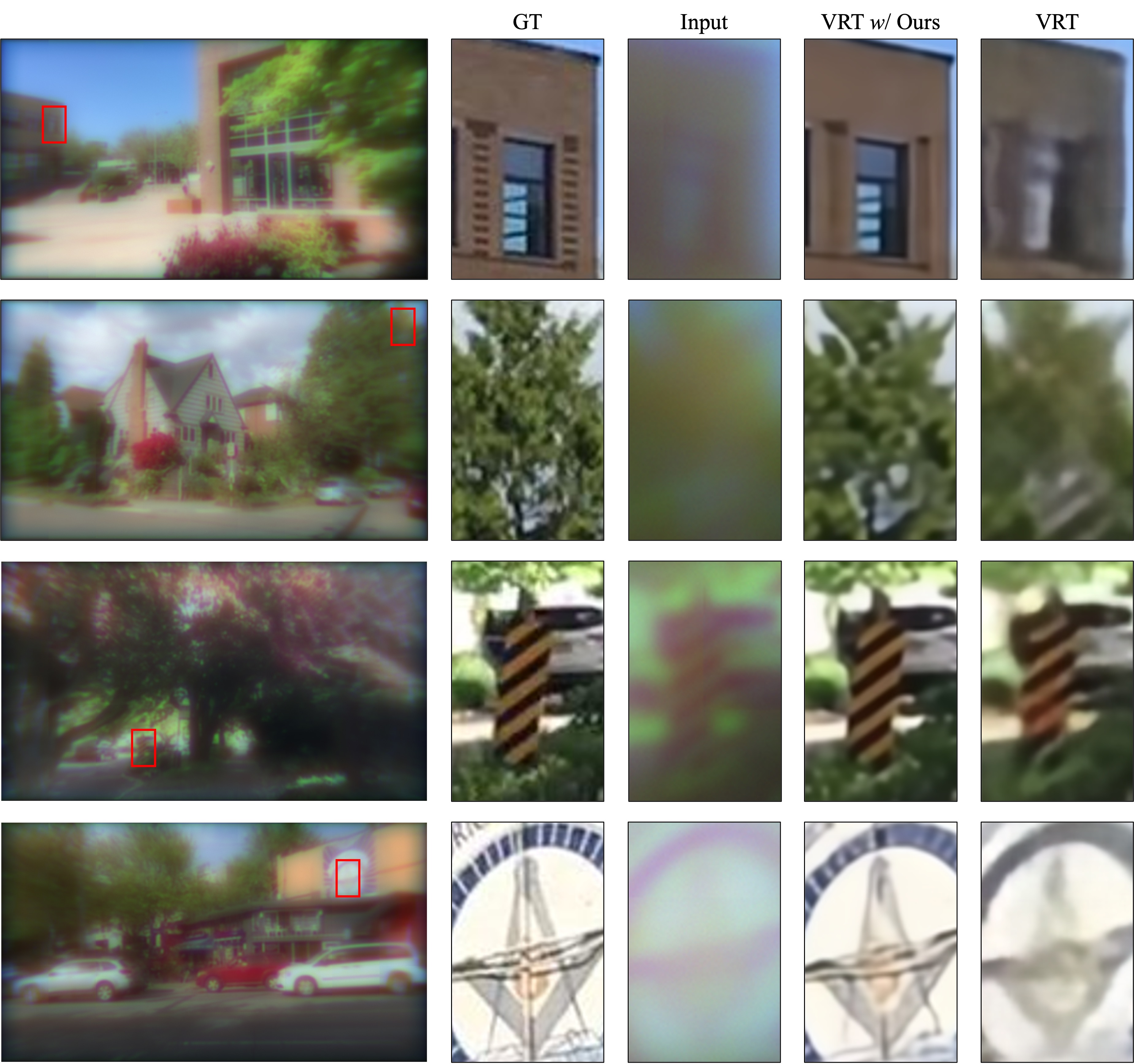}
\end{center}
\vspace{-0.5cm}
   \caption{Additional qualitative results on the aberrated DVD~\cite{dvd} dataset.}
\vspace{-0.2cm}
\label{fig:video_qualitative_supp}
\end{figure*}

\cref{fig:video_qualitative_supp} shows additional qualitative results on the aberrated DVD~\cite{dvd} dataset. Similar to~\cref{fig:quali_video} in the main paper, the proposed method could robustly address aberrations in the video and produce aberration corrected representations.

\subsection{3D Reconstruction with Aberrated Images}
\noindent\textbf{Implementation Details}
We set the loss weighting hyperparameter $\lambda = 1.0$ for the Tanks\&Temples~\cite{tanks} datasets. For the LLFF~\cite{llff} dataset, we set $\lambda=0.75$ and inactivated periodically resetting opacity of 3D Gaussians to preserve more points in the forward-facing dataset~\cite{deb3dgs}. We use default hyperparameters including learning rate, training iterations and 3D Gaussian densification parameters from the original paper~\cite{3dgs} for our experiments.

\begin{figure*}
\begin{center}
\includegraphics[width=0.95\linewidth]{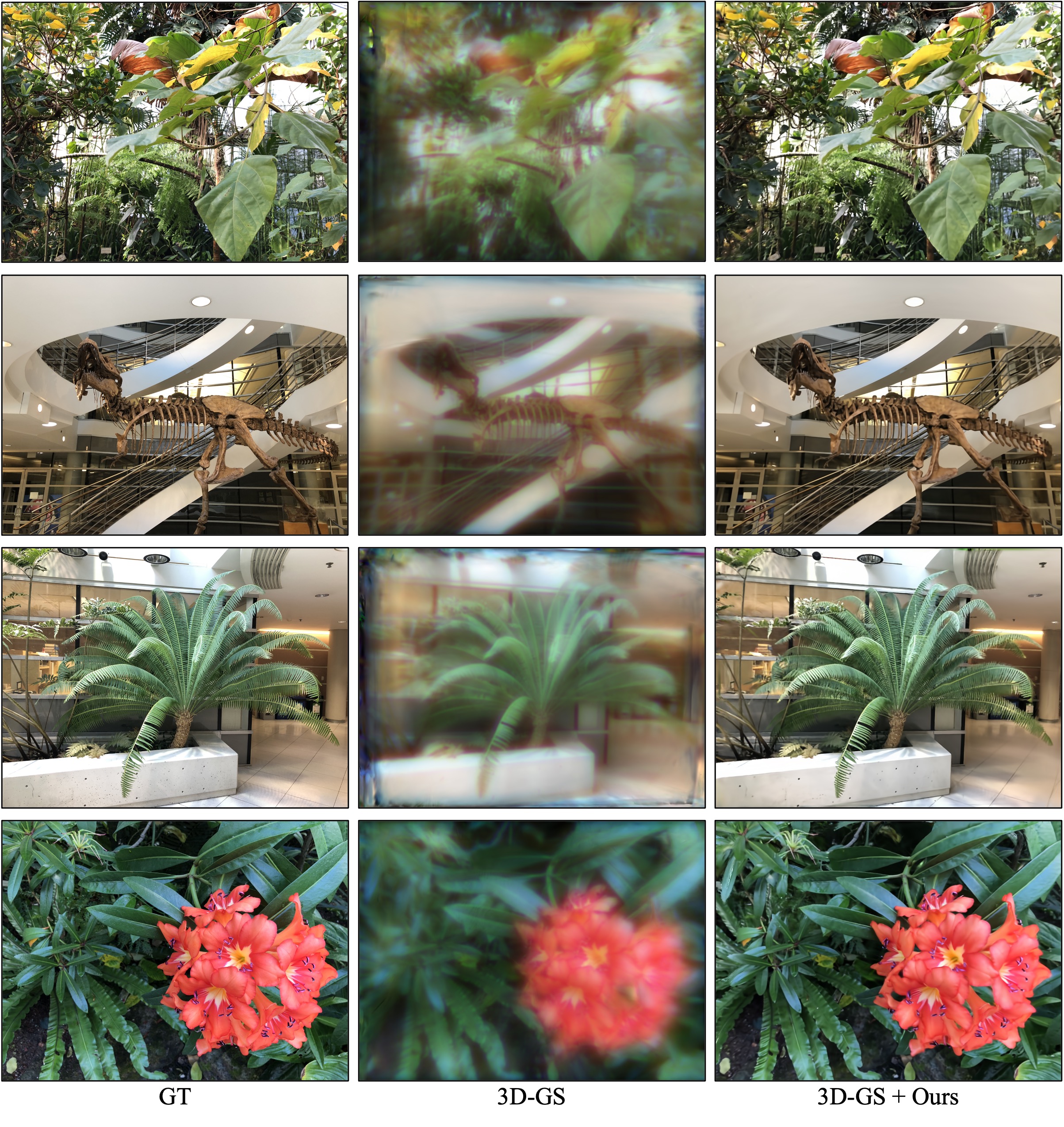}
\end{center}
\vspace{-0.8cm}
   \caption{Qualitative results on the aberrated LLFF~\cite{tanks} dataset.}
\vspace{-0.2cm}
\label{fig:quali_nvs_llff}
\end{figure*}

\begin{figure*}
\begin{center}
\includegraphics[width=0.95\linewidth]{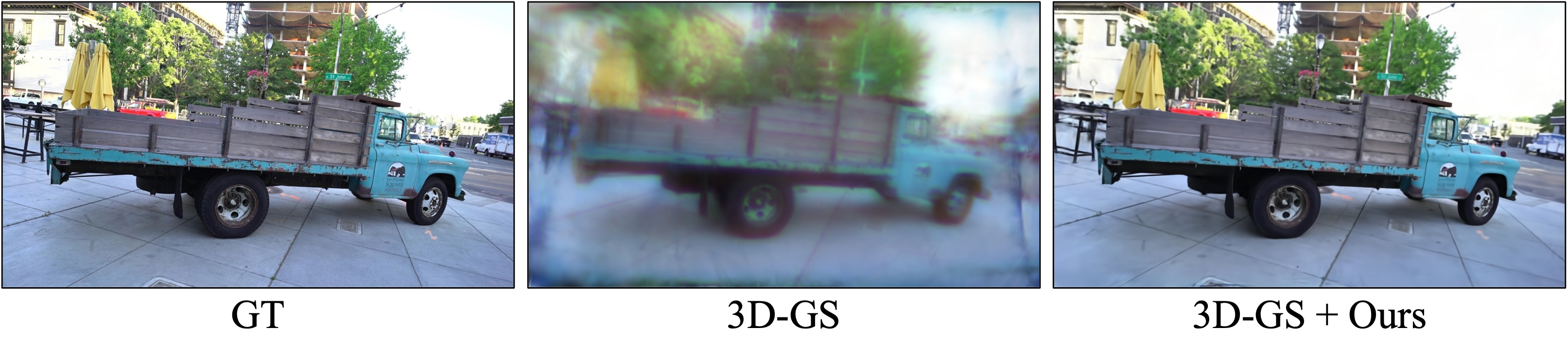}
\end{center}
\vspace{-0.8cm}
   \caption{Qualitative results on the aberrated Tanks\&Temples~\cite{tanks} dataset.}
% \vspace{-0.2cm}
\label{fig:quali_nvs_tnt}
\end{figure*}

\begin{figure*}
\begin{center}
\includegraphics[width=0.95\linewidth]{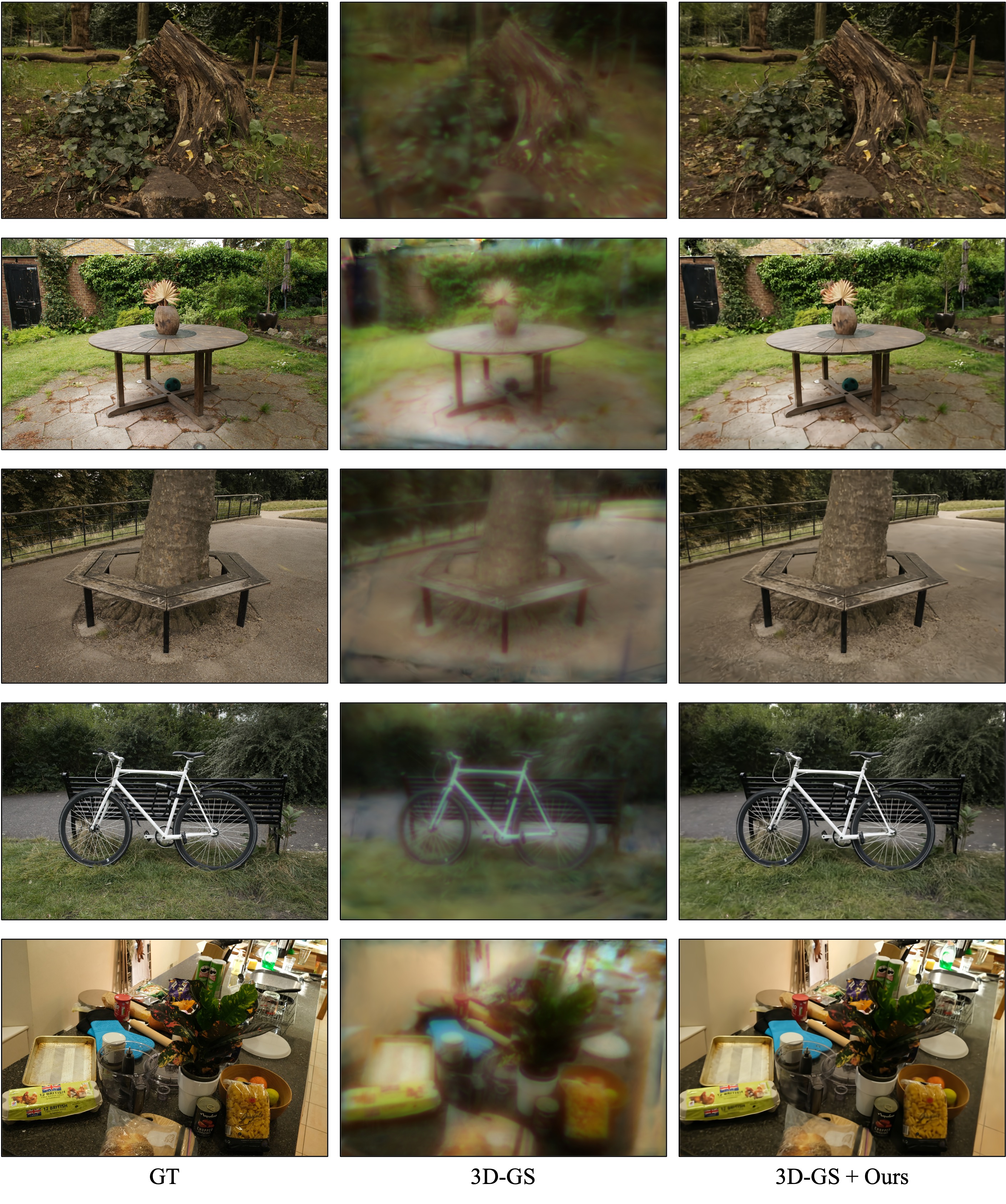}
\end{center}
\vspace{-0.8cm}
   \caption{Qualitative results on the aberrated Mip-NeRF 360~\cite{mip360} dataset.}
% \vspace{-0.2cm}
\label{fig:quali_nvs_360}
\end{figure*}

\noindent\textbf{Results} We provide extra qualitative results on the LLFF~\cite{llff} and Tanks\&Temples~\cite{tanks} datasets in~\cref{fig:quali_nvs_llff} and~\cref{fig:quali_nvs_tnt}.
We further conducted experiments on Mip-NeRF 360~\cite{mip360} dataset with spatially varying aberration, where we set $\lambda = 1.0$. As described in~\cref{tab:quant_nvs_360}, the proposed training pipeline that embed our pre-trained restoration network into 3D-GS could improve the scene reconstruction quality through additional multi-view consistency guidance. \cref{fig:quali_nvs_360} shows that 3D-GS could not model the spatially varying aberration of metalens and failed to produce clean representations. On the other hands, the proposed method enables aberration modeling, demonstrating the capability of clean 3D reconstruction from the set of multi-view aberrated images of our approach.

\begin{table}[]
    \centering
    \begin{tabular}{c|ccc}
    \hline
    Method & PSNR$\uparrow$ & SSIM$\uparrow$ & LPIPS$\downarrow$\\
    \hline
    \hline
    3D-GS~\cite{3dgs} & 19.58 & 0.3857 & 0.8166 \\
    3D-GS \textit{w/} Ours & 25.88 & 0.6669 & 0.3696 \\
    3D-GS + Ours & \textbf{25.94} & \textbf{0.6676} & \textbf{0.3693} \\
    \hline
    \end{tabular}
    \caption{Quantitative results on the Mip-NeRF 360~\cite{mip360} dataset with spatially varying aberration.}
    \label{tab:quant_nvs_360}
\end{table}

\section{Limitations and Future Work}
We fabricated a metalens using the phase profile of~\cite{tseng2021neural} and captured images projected on the display, following its optical setup. However, capturing actual scenes can introduce various physical variables, such as multiple light sources and noises, which can undermine the quality of the images, but the current optical system is not well-suited for in-the-wild imaging. To address this limitation, we are designing a new lens optimized for real world applications, and propose a more robust and effective restoration network capable of handling metalens-captured scenes as a future work.

\end{document}